\documentclass[final]{elsarticle}

\usepackage{lineno,hyperref}
\modulolinenumbers[5]

\usepackage[margin=0.8in]{geometry}

\usepackage{natbib}
\usepackage{graphicx}
\usepackage{float}
\usepackage{bbold}
\usepackage{amsmath}
\usepackage{amssymb}
\DeclareMathOperator{\tr}{tr}
\usepackage{amsfonts}
\usepackage{amssymb}
\usepackage{mathrsfs}
\usepackage{mattens}
\usepackage{mhchem}
\usepackage{amsthm}
\usepackage{upgreek}
\usepackage{textcomp}
\usepackage{placeins}
\usepackage{gensymb}
\usepackage{subcaption}
\usepackage{textcomp}

\usepackage{soul}
\usepackage{todonotes}

\usepackage{xcolor}
\usepackage{framed}
\colorlet{shadecolor}{blue!20}

\usepackage{listings}

\let\vec\mathbf
\let\mat\boldsymbol

\journal{Journal of the Mechanics and Physics of Solids}

\begin{document}

\begin{frontmatter}

\title{Contact phase-field modeling for chemo-mechanical degradation processes.\\
Part I: Theoretical foundations.}

\author[mymainaddress]{A. Gu\'evel}
\corref{mycorrespondingauthor}
\cortext[mycorrespondingauthor]{Corresponding author}
\ead{alexandre.guevel@duke.edu}

\author[mymainaddress]{H. Rattez}

\author[mysecondaryaddress]{S. Alevizos}

\author[mymainaddress]{E. Veveakis}

\address[mymainaddress]{Duke University, School of Civil and Environmental Engineering, USA}
\address[mysecondaryaddress]{National Technical University of Athens, Department of Mechanics, Greece}

\begin{abstract}
As phase-field modeling (PFM) is booming across various disciplines and has been proven fitted for numerically modeling interfacial problems, we aim at taking a step back to revisit its fundamental validity, in the light of non-equilibrium thermodynamics. For that, a general contact thermodynamics (CT) framework is derived from contact geometry, based on the maximum dissipation principle (MaxDP), thus extending Gibbs\textquotesingle \ seminal geometrical representation of thermostatics. Combining CT and micro-force balance, the gradient flow equation usually derived for PFM from the variational formulation can be written as generalized relaxation equations. The obtained viscous Allen-Cahn equation allows both the PFM kinematic degrees of freedom, the order parameter and its gradient, to be fully dissipative. The model is also extended to a double PFM, in order to include chemo-mechanical coupling, corresponding respectively to endothermic and exothermic processes and thus leading to a phase change bidirectionality. This contact PFM (CPFM) will be applied in the second part of this work to irregular microstructures like geomaterials, valid for porous media in general, with a focus on pressure solution.

\end{abstract}


\end{frontmatter}


\section{Introduction} 


\subsection{Phase-field modeling}

PFM has been shown to be a fitted numerical tool to model interfacial problems. By smoothing the physical sharp interface, it avoids the mathematically and numerically tedious tracking of the interface. Since its theoretical foundations in the 70s, there is little need to show the apparent success of  the plethora of its numerical applications over the last two decades. Following the seminal works of Fix and Langer \cite{Fix1983} for first order liquid to solid phase transition and Fried and Gurtin for solid-liquid and solid-solid transitions \cite{Fried1993, Fried1994}, PFM has been applied to a wide range of fields. In material sciences, PFM, along with databases like CALPHAD, provides satisfying quantitative description of multi-component alloys \cite{Zhang2012}, and specifically interfacial instabilities like dentritic growth \cite{Kim1999}. Precipitation and dissolution models \cite{Xu2014} are good example of the multiphysic flexibility of PFM. The recent applications to unsaturated media are of prime importance to the field of geophysics \cite{Cueto-Felgueroso2009a, Sciarra2016a}. In biology, given the ubiquity of interfaces processes in the human body, it is not surprising that PFM has gained importance as well, such as for modeling tumours \cite{Cristini2009,ODEN2010,Garcke2015} or vesicle membranes \cite{Du2004}. The sky is not the limit for PFM as exotic applications like modeling Saturn's rings \cite{Tremaine2003} can be pointed out. Essentially, PFM could in theory model any processes with interfaces. But more than pushing further the extent of those applications, our will is, in a first step (back), to shed light on the fundamental validity of PFM in the context of non-equilibrium thermodynamics.\\

The seminal theories of PFM, based on the concepts of order parameter of Landau {\cite{Landau1937}} and diffuse interface of Cahn {\cite{Cahn1958}}, have been developed chiefly within Carnegie Mellon University. First, Langer in 1978 in lecture notes, whose results were first published by Fix in 1982 {\cite{Fix1983}}, defined PFM by a gradient flow equation:

\begin{equation} \label{variational_PFM}
\tau\dot{\phi} = -\frac{\delta F}{\delta \phi}
\end{equation}

With {$\tau$} the relaxation time to equilibrium, {$F$} the free energy (integral) functional and {$\phi$} the order parameter. This formulation is equivalent to minimizing the free energy of the system, or in other words to have the system relax as fast as possible to equilibrium, modulo a relaxation time. This variational formulation is the framework mainly used nowadays. Then, Fried and Gurtin derived the PFM equations within a framework of configurational forces and continuum thermodynamics {\cite{Fried1993}}. Both derivations allowing non-equilibrium processes to a certain extent, it seems that the main constitutive assumption boils down to respectively the minimization of the free energy or the MaxDP, as we will discuss further (cf part \ref{comparison with variational formulation}). Such driving assumption is indispensable to prescribe the behaviour of a system out of equilibrium, inasmuch as the second law of thermodynamics only provides a necessary but not sufficient condition then. 

However, we argue through our derivation that a term is missing, the Laplacian rate $\Delta\dot\phi$. Indeed, it is crucial to keep in mind that by adding the gradient of the order parameter in the system's state variables, PFM is a higher order theory (gradient theory) of the sharp interface theories. As such, if we consider (fully) dissipative structures, i.e. the type of structures that allow systems self-organization when far enough from equilibrium {\cite{Prigogine1978}}, it is as much important to allow the kinematics of the order parameter to be dissipative as to allow that of its gradient. Indeed, the description of the dynamics of any interface should allow two degrees of freedom, by mathematical definition of a surface. In the case of PFM, those degrees of freedom correspond to the normal variations of the interface ($\dot\phi$) and the variations of its orientation ($\Delta\dot\phi$) (cf eq.\ref{curvature}). Furthermore, since we derive PFM from the MaxDP, it is consistent that all the state variables end up being dissipative.  Hence, it seems that the presence of the latter term is as important as the former one. This has been derived first by Gurtin by considering the microstress dissipative, i.e. by adding $\nabla\dot\phi$ in the constitutive variables. This flexibility of the theory of Fried and Gurtin alleviates the limitations of the variational formulation for fully dissipative processes, given that the latter "limits the manner in which rate terms can enter the basic equations" {\cite{Fried1993}}. We use yet a different approach regarding the inclusion of the rate terms. We aim at going further and attempting to justify that a PFM derivation should intrinsically contain the viscous term $\Delta\dot{\phi}$. We do not object nevertheless that the influence of this term may be negligible in certain cases, namely when the change of curvature is negligible. Our main motivation here is to rederive in the most general way the PFM equations, starting explicitly from the main assumptions, without making use of the variational formulation. For that, we shall not restrain in any way the system to be close or not to equilibrium, and we will formally apply the MaxDP in the most general context, that of CT, before deriving the PFM equation. To deal with distance (from equilibrium) and measurement (maximization), the contact geometry is to be endowed with a metric. It is only then that we may appeal to the fundamental laws, thus ensuring a clear separation of the balance laws from  the constitutive equations, as recommended by Gurtin {\cite{Gurtin1996}}.


\subsection{Maximum dissipation principle} 

\subsubsection{Primordiality of a general constitutive assumption}

A general constitutive assumption is primordial for non-equilibrium thermodynamics processes and to ensure a clear distinction between the fundamental laws of nature and the assumed constitutive assumptions.

Firstly, conventional thermodynamics, i.e. thermostatics, provides a framework to define a system's equilibrium states but only a guide to non-equilibrium processes. The first law describes the equilibrium states through Gibbs' tangent planes representation \cite{Gibbs1873}, formalized later in Legendre submanifolds in the contact geometry representation \cite{Hermann1973}. The second law guides the system out of equilibrium by providing the admissible processes \cite{Haslach1997, Haslach2011}, providing therefore solely a necessary condition, obtained via the Coleman-Noll procedure \cite{Coleman1963}. To fully prescribe a non-equilibrium process, there is no choice but choosing one of the admissible constitutive assumptions; the choice of the latter is another question. This complementary principle takes all its sense when remembering that thermodynamics has been initially built as a black box description. It shall provide a  complementary support for the description for the internal structure of the system, to describe the internal fluxes (self-organization) responding to the environment's external solicitations. This point of view is fully developed in Bejan's constructal theory \cite{Bejan2016}, first stated in \cite{Bejan1997}: "For a finite-size flow system to persist in time (to live), it must evolve in such a way that it provides easier access to the imposed (global) currents that flow through it".

Secondly, we advocate a clear distinction between the fundamental laws and the constitutive assumptions, following the initiative of rational thermodynamics and in particular the work of Fried and Gurtin (see e.g. \cite{Fried1993, Fried1994, Gurtin1996}). For that, the constitutive assumption should be independent from the fundamental laws and thus as general as possible. That avoids resorting to ad-hoc assumptions when needed in the course of a model's derivation. In that sense, a principle prescribing a quantity as general as the dissipation appears to us as a good candidate. All the constitutive relations then obtained in the model will be knowingly and consistently due to such founding principle.


\subsubsection{The choice of the MaxDP}

A wide discussion is deserved on the choice of MaxDP, with support from various fields. As far as we are concerned in mechanics, it is predominantly used for plasticity models, as already referenced in \cite{Haslach2011} for instance. As referenced  by Lubliner in \cite{Lubliner1984}, the use of the MaxDP (called in this particular case principle of maximum plastic dissipation) originated with von Mises (1928) and Hill (1948), restricted to rigid plasticity. Mandel (1964) (ref. in \cite{Lubliner1984}) extended it to elasto-plasticity and Lubliner to the case of large deformations \cite{Lubliner1984}. Modern plasticity and the widely used normality rule thus stem from the MaxDP. Essentially it is the particular case, with regard to the general model developed in the present work, where the plastic strain is a state variable and the stress is the associated control variable. It must be emphasized that the plastic strain can be considered as a state variable only because the plastic law is derived with respect to a reference stress within or on the yield surface. 
A similar approach is due to Ziegler who generalized Onsager's principle to nonlinear phenomenological laws \cite{Ziegler1958}. It is later coined as Maximum Entropy Production Principle (MaxEPP). This yields Ziegler's orthogonality rule, generalizing the maximum plastic dissipation principle.  Even though this rule seems fairly general, it is still taken with care and rightly emphasized more as a "classifying hypothesis" \cite{Houlsby2000} than a general rule. We also do stress that the choice of the MaxDP is a mere constitutive assumption and debatable, at least until further progress.

Notwithstanding, we argue that beyond the apparent restriction of the MaxDP and its arbitrary choice, in the general context of CT it seems to bear a broader constitutive meaning. As discussed in parts \ref{Clarifications on thermodynamics} and \ref{Break/Engine behavior}, our formulation could be a formalization of the constructal law, which happens to maximize the system's dissipation; the latter does not have to be the starting point. Another supporting direction is the one taken by modern statistical non-equilibrium thermodynamics, allegedly led by England's "adaptative dissipation" theory \cite{England2013,England2015}. This theory describes systems at the statistical scale (based on the fluctuation theorem) following the direction of maximum dissipation when disturbed by micro-fluctuations. Note that as pointed out by Prigogine \cite{Prigogine1996}, those fluctuations become preponderant far enough from equilibrium, and the smaller the scale the more fluctuations. England thus applied consistently the theory of adaptative dissipation to the emergence of life from the molecules scale, where natural systems seemingly at rest can be triggered by micro-fluctuations. Another parallel can be made with Deep Learning, where the stochastic gradient (or steepest) descent algorithm (SGD) guides the system to follow the information gradient, so that the system reaches equilibrium (meaning in that context coincidence of the system's response with the observations) as fast as possible, given the available information given by the environment. It is our understanding that CT is based on similar considerations, except that the discrete information from the environment is transformed into continuous processes. Indeed, we describe in the present work, stemming from contact geometry, a discrete contact structure tangenting the space of processes, thus ensuring the "contact" between the system and its environment (cf fig.\ref{fig:sketch_contact_thermodynamics}). The contact structure's information is translated into processes via the MaxDP, pushing the system to reach equilibrium as fast as possible (we prove it mathematically in part \ref{Break/Engine behavior}). All in all, we find comfort in assuming the MaxDP in the sense that it seems to build on an overarching principle, seemingly related to Darwinism. Indeed, as for Bejan's constructal theory and England's adaptative dissipation, the system (optimally) evolves to adapt to its (evolving) environment and the fittest survives.

 \subsubsection{Clarifications on thermodynamics} \label{Clarifications on thermodynamics}

Thermodynamics has been evolving ever since its empirical foundations. In order to formalize its laws and apply them more broadly, especially to non-equilibrium processes, a reformulation is required. This is most relevant to biology and patter formation in general, for an increase in "order' in the system may seem to violate the second law. The confusion around patterns formation has even led Shrodinger to preconize a separation of the description of the animate and the inanimate. This false paradox has since been cleared out by considering out-of-equilibrium processes and realizing that "entropy" is not synonym of "disorder" (see \cite{Martyushev2013} for a thorough account of those misconceptions). We intend to contribute further to the clarification of such misconceptions.

First of all, it is primordial to be clear about the thermodynamic foundations, the first and second laws. While the first law is merely the energy balance (which definition can be arranged when needed), the second law remains rather unclear insasmuch as a plethora of statements exists. As Truesdell wrote in \cite{Truesdell1969}, there seems to be as many thermodynamic theories as thermodynamicists. More recently, a comprehensive study on the second law and its challenges referenced 21 one different statements, as well as 21 different entropy varieties (which is most likely a non-exhaustive list). The most troubling statement but also the most popular one is that entropy of an isolated system will never decrease. Oftentimes, the simplification by "disorder always increases" causes confusion, especially when dealing with systems' self-organization. A good start is, rather than creating "new basic physical axioms", to carry out a "rational classification of the variety compatible with the previously knows axioms" \cite{Truesdell1969}. Truesdell and his coworkers such as Coleman and Noll dramatically rationalized the formulation of thermodynamics \cite{Truesdell1969} and constructed their continuum thermodynamics framework to be compatible with modern continuum mechanics. In particular, the so-called Clausius-Duhem Inequality (CDI), formalized in \cite{Truesdell1960}, is a consequence of the second law and considered as the second law in continuum thermodynamics:

\begin{equation} \label{CDI}
\dot{\eta} \geq -\nabla.{\vec{q}/\theta} + r/\theta
\end{equation}

Where $\eta$ is the entropy, $\vec{q}$ the heat flux, $\theta$ the temperature and $r$ the heat supply \footnote{here and in all this paper, we note in bold characters the vectors and tensors}. It shall be kept in mind that it is a practical restriction of the second law; it does not always imply the latter but appears to be usually valid. Let us now write the first law (energy balance):

\begin{equation} \label{first law}
\dot{e}=-\dot{W}-\nabla.\vec{q}+r
\end{equation}

Where $e$ is the internal energy and $\dot{W}=\vec{y}.\vec{\dot{x}}$ is the power expenditure (the "useless" work spent by the system), $\vec{x}$ and $\vec{y}$ are respectively the vectors of the state and control variables (for instance strain and stress respectively). We use the convention that it is positive for endothermic processes (the system receives energy from its environment) and negative for exothermic processes (the system gives energy from its environment). For instance, for the mechanical dissipation we will choose $y=-\sigma$ and then the corresponding power expenditure will be negative (exothermic). Combining eq.\ref{CDI}, eq.\ref{first law} and the definition of the free energy $\Psi=e-\theta\eta$ yields the dissipation inequality:

\begin{equation} \label{DI}
\dot\Psi + \dot{W} + \eta\dot{\theta} + \vec{q}.\frac{\nabla \theta}{\theta}=-D \leq 0
\end{equation}

This provides an (almost) necessary condition for admissible processes, which can be applied to restrict the constitutive laws of a system via the Coleman-Noll procedure \cite{Coleman1963}. Note that our power expenditure term characterizes the dissipation due to any state variables. In the rest of this paper, we will consider for simplicity isothermal conditions, hence we just consider the two thermodynamic laws combined in one expression:

\begin{equation} \label{isothermal 2 laws}
\begin{cases}
D=\theta\dot\eta \\
D \geq 0
\end{cases}
\end{equation}

With $D=-\dot\Psi^*-\dot{W}$. This appears as the mere expression $D=\theta\dot\eta \geq 0$. However we will see that in the following non-equilibrium thermodynamics framework, that, paramountly, the 1st law only applies at equilibrium (on certain submanifolds of the thermodynamic space) whereas the second law holds everywhere, including out of equilibrium. This thusly corroborates the limitation of the entropy-always-increasing statement of the second law to isolated systems. Indeed, as discussed in part \ref{Contact thermodynamics}, equilibrium means in the context of CT no "contact" between the system and its environment. In that sense, the interaction between the system and its environment should be at the heart of thermodynamics. When the system is in dynamic contact with its environment (not isolated), its entropy does not have to increase so that it can self-organize; but for that, it has to be out of equilibrium, whence the necessity to study the associated processes in a non-equilibrium thermodynamics framework.

The second point of confusion that attracted our attention regards extremum principles formulations, the MinEPP (Prigogine) and MaxEPP (Ziegler), which can be easily confused to contradict each other. As explained in details in \cite{Martyushev2013}, the problem is that they concern different time scales, giving raise to a "hierarchy of processes": on short time scales (when the thermodynamic forces can be considered fixed), the system adjusts its fluxes so that the MaxEPP prevails, whereas on much longer time scales the MinEPP prevails. Martyushev attributes the smaller time scale to the diffusion at the microscale (molecular) and the bigger scale to that where the forces evolve. We attribute this hierarchy MaxEPP/MinEPP respectively to the two ideal brake/engine behaviors (see \ref{Break/Engine behavior}), which is reconciled in our MaxDP. We can also note that England's "dissipative adaptation" theory, based on dissipation maximization, can imply both MaxEPP and MinEPP \cite{England2013}.\\

All in all, we hope that working in a non-equilibrium thermodynamics framework will allow to shed light on such hindrances that have been curbing the progress of thermodynamics.



\subsection{Non-equilibrium thermodynamics} 

Finally, we may introduce the context of non-equilibrium thermodynamics before constructing the model within its framework in the next part. A thorough summary is provided in Haslach's book \cite{Haslach2011}. From the first empirical deductions of Carnot in the $19^{th}$ century, through the subsequent diverse formulations of the laws of thermodynamics, the thermostatics theory of Gibbs, to the numerous attempts to define and model non-equilibrium thermodynamics over the $20^{th}$ century until now, thermodynamics has been one of the most challenging fields to develop and still remains rather unfinished. All the more so as quantum and statistical mechanics have been challenging its validity at the lowest scales. The link between those two apparently contradicting worlds may be found when the system is far from equilibrium. According to Prigogine, while the fluctuations of statistical mechanics are damped near equilibrium, they may become preponderant far from it, and actually explain the appearance of nonlinear dynamic phenomena like bifurcations \cite{Prigogine1996}, or even life as seen in recents works as England's theory \cite{England2013}. Prigogine argues as well that “near-equilibrium laws of nature are universal, but far from equilibrium, they become mechanism dependent”  \cite{Prigogine1996}. Prigogine's questioning on far-from-equilibrium processes, and the problem of the arrow of time in particular, induced a major paradigm shift in thermodynamics, leading to considering the discipline as a holistic framework to model nature. Thus thermodynamic irreversibility found all of its meaning and importance at the root of modern thermodynamics and led "from trajectories to processes" \cite{Prigogine1978}. In that sense, we aim at developing a seminal thermodynamic description of processes, that we call GRE. We also discuss in \ref{source_of_irreversibility} the implications of our theoretical developments in the light of Progine's ideas. We will see that mechanism-dependent a priori means rate-dependent. From there comes the necessity of defining a metric to define how far a system is from equilibrium. All those hints naturally point towards using the intrinsic geometrical framework of thermodynamics, contact geometry, as stated by Arnold and first employed by Gibbs \cite{Gibbs1873}: "Every mathematician knows that it is impossible to understand any elementary course in thermodynamics. The reason is that the thermodynamics is based - as Gibbs has explicitly proclaimed - on a rather complicated mathematical theory, on the contact geometry." \cite{Arnold1990}.

\section{Contact thermodynamics} \label{Contact thermodynamics}

Building on the previous ideas, we naturally tend towards choosing a geometrical representation of thermodynamics, and its natural geometry is contact geometry \cite{Hermann1973,Arnold1990, Haslach1997}.

\subsection{Construction}

The maximum dissipation non-equilibrium thermodynamic model, developed by Haslach initially in \cite{Haslach1997} and thoroughly in \cite{Haslach2011}, is largely the inspiration for the thermodynamic framework of our model. The formalization of thermodynamics under contact geometry has been initiated by Hermann in 1973 \cite{Hermann1973}. A clear outlook of the mathematical structure of thermodynamics is given as well by Salamon et al. in \cite{Salamon2006} for instance. This framework will be referred to as "contact thermodynamics" (CT). As claimed by Arnold \cite{Arnold1990}, contact geometry provides a mathematical structure to thermodynamics, which is essential to derive a sound and clear model, especially when it comes to non-equilibrium thermodynamics. CT generalizes Gibbs' seminal idea of representing a system at equilibrium geometrically with the graph of a thermostatic energy function \cite{Gibbs1873}. This is achieved practically by generalizing this energy function, following Haslach's model \cite{Haslach1997, Haslach2011}.

We thus represent the thermodynamic phase space (TPS) as a contact manifold $(\mathcal{M},\omega)$, where $M$ is a smooth manifold of dimension $2n+1$ and $\omega$ is a contact form, called the Gibbs form in the present thermodynamics context (cf \ref{Mathematical toolbox} for a mathematical background). The natural integer $n$ is the number of degrees of freedom of the thermodynamic system, i.e. the number of state variables by which it can be adequately modeled. A thermodynamic system is geometrically represented in $(\mathcal{M},\omega)$ by a codimension one submanifold (of dimension $2n$), a fortiori a symplectic manifold (cf \ref{Mathematical toolbox}), which is locally the graph of the generalized energy function, noted $\Psi^*$. This symplectic manifold enables to naturally identify the thermodynamic conjugate pairs $(x_1,...,x_n,y_1,...,y_n)$ as a set of coordinates, where the $x_i$ are the state variables and the $y_i$ the associated control variables. A physical thermodynamic system is thus fully determined by the graph of $\Psi^*$ equipped with the thermodynamic pairs $(x_i,y_i)$. $(\Psi^*,x_1,...,x_n,y_1,...,y_n)$ forms a set of coordinates for the contact manifold $(\mathcal{M},\omega)$ called the contact coordinates. Non-equilibrium processes are paths on the graph of the generalized energy function, guided by the tangential action of the Gibbs contact form $\omega$, which guarantees the Clausius-Duhem inequality for admissible processes and acts as a measure of the dissipation, as shown below. When the Gibbs form vanishes, the system reaches a Legendre submanifold $\tilde{\mathcal{M}}$ (of dimension $2n$), geometrical representation of thermodynamic equilibrium in the sense that the system runs out of energy to dissipate. Note that this equilibrium is usually dynamic (time-dependent) since the Legendre submanifolds are time dependent when the control variables are. Any points outside the Legendre submanifolds represent non-equilibrium states.\\

 Following Salamon's more intuitive description \cite{Salamon2006},  the $x_i$ describe the system, the $y_i$ describe the environment, and $\Psi^*$ ensures the contact between the $x_i$ and the $y_i$, i.e. between the system and its environment. The way the contact is performed, i.e. the system's dynamics, is prescribed by the contact form $\omega$ ($\omega$ is a function of $x_i$, $y_i$ and $\Psi^*$). Thus the term "contact", more than a geometrical abstract denomination, describes in thermodynamics the actual contact between a system and its environment. It is interesting to remember that Gibbs' seminal representation of (contact) thermodynamics may have been inspired by his mechanical engineering background and his thesis on the ideal shape of spur gearing. The first representation of CT was a sculpture by Maxwell in 1874 inspired by Gibbs' graphs in \cite{Gibbs1873}:

\begin{figure}[h!]
\centering
\includegraphics[scale=0.5]{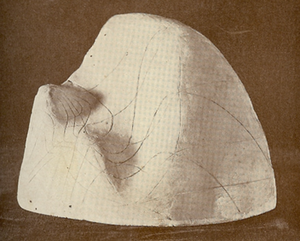}
\caption{Photo of Maxwell's plaster model, called "Gibbs thermodmamics surface" (taken by James Pickands II, and published in 1942)}
\label{fig:maxwell}
\end{figure}

The Gibbs form is crucial as it is the driving quantity for the system out of equilibrium in this geometrical context and it actually embodies the two first laws of thermodynamics by definition, as shown in the next part. The requirement for its tangential action to be non-positive $\omega(t_p) \leq 0$ is equivalent to the Clausius-Duhem inequality \cite{Haslach2011, Salamon2006}, i.e. the second law . When the system reaches equilibrium, $\omega=0$ yields the first law \cite{Gibbs1873,Arnold1990, Salamon2006,Haslach2011}. This will be further clarified in part \ref{Recovering the first and second laws}.

As per Haslach's model \cite{Haslach1997}, we extend the definition of the system's thermostatic energy function $\Psi$ to non-equilibrium thermodynamics by defining a generalized energy function $\Psi^*$ on the symplectic manifold defined by the state and control variables:\\

\begin{math}
\begin{array}{l|rcl}
\Psi^* : & \mathbb{R}^{2n} & \longrightarrow & \mathbb{R} \\
    & (x_i,y_i) & \longmapsto & \Psi^*(x_i,y_i) 
\end{array}
\end{math}
\\

We thus start from the system's equilibrium description, its equations of state (only valid at equilibrium), with $\hat\Psi$ the system's equilibrium energy:

\begin{equation} \label{EOS}
x_i \equiv \frac{\partial\hat\Psi}{\partial{y_i}}  \qquad   \forall i \in [\![1,n]\!] \qquad on \ \tilde{\mathcal{M}}
\end{equation}

This is usually the way a thermodynamic (or rather thermostatic) system is defined (cf \cite{Houlsby2000} e.g.). Equivalently, noting $\Psi$ the Legendre transform of $\hat\Psi$:

\begin{equation} \label{EOSbis}
y_i \equiv -\frac{\partial\Psi}{\partial{x_i}}  \qquad   \forall i \in [\![1,n]\!] \qquad on \ \tilde{\mathcal{M}}
\end{equation}

We now define the generalized energy function $\Psi^*$ with the two requirements here below. The main idea is that $\Psi^*$ Legendre conjugates the system's thermodynamic pairs $(\vec{x},\vec{y})$ out of equilibrium and $\Psi$ does so only at equilibrium ($\Psi^*$ and $\Psi$ coincide at equilibrium). $\Psi^*$ thus ensures the "contact" between the system ($\vec{x}$) and its environment ($\vec{y}$) beyond equilibrium.\\

1) $\Psi^*$ shall Legendre-conjugate the state ($\vec{x}$) and control variables ($\vec{y}$):

\begin{equation} \label{Legendre_conjugation}
\frac{\partial\Psi^*}{\partial{y_i}} \equiv x_i   \qquad   \forall i \in [\![1,n]\!] 
\end{equation}

2) The affinities shall vanish on the system's Legendre submanifolds $\tilde{\mathcal{M}}$:

\begin{equation} \label{affinity_definition}
X_i \equiv \frac{\partial\Psi^*}{\partial{x_i}} = 0  \qquad   \forall i \in [\![1,n]\!] \qquad on \ \tilde{\mathcal{M}}
\end{equation}

We note $\Psi=\Psi^*_{\restriction{\tilde{\mathcal{M}}}}$. Integrating eq.\ref{Legendre_conjugation} yields: 

\begin{equation} \label{generalized_free_energy}
\Psi^*(x_i,y_i) = \Psi(x_i) + \sum_{i=1}^{n} x_iy_i
\end{equation}

One can check that the equations of state (eq.\ref{EOSbis}) are recovered on $\tilde{\mathcal{M}}$ (where eq.\ref{affinity_definition} holds).

Note that if the system is not forced, i.e. if the control variables are constant in time, $\tilde{\mathcal{M}}$ corresponds to pure equilibrium. If it is forced, $\tilde{\mathcal{M}}$ is time dependent and  we may speak of a dynamic equilibrium. The equilibrium is stable if the corresponding Hessian $\frac{\partial^2\Psi}{\partial{\vec{x}}^2}$ is positive definite \cite{Haslach1997,Haslach2011}. We corroborate this criterion by rewriting our GRE in part \ref{Break/Engine behavior} (cf eq.\ref{GRE_exponential_relaxation}).

Thus, once the $n$ thermodynamic pairs $(x_i,y_i)$ ($\forall i \in [\![1,n]\!]$) are chosen, defining the generalized energy function boils down to defining the equilibrium energy function, which is the only part one can access to or measure anyway. Now that the system's variables and energy are defined (i.e. the contact between the system and its environment is modeled), it remains to prescribe its dynamics, i.e. its behaviour outside equilibrium. For that, we define the Gibbs form with the contact coordinates ($\Psi^*$, $x_i$, $y_i$), in the Darboux canonical form (any other choice of coordinates is reducible to this one):

\begin{equation} \label{contact_form}
\omega = d\Psi^* - \sum_{i=1}^{n} x_idy_i
\end{equation}

This result called the "relative Darboux theorem for contact forms" and derived by Arnold in \cite{Arnold1990b} comes from the Darboux theorem.
$\Psi^*$ is said to be the potential function of the canonical set $(x_i,Y_i)$ \cite{Hermann1973}. A potential function of another canonical set is then a Legendre transform of $\Psi^*$.

Its action on the tangent vector \cite{Haslach2011} reads (cf \ref{Tangential action of the Gibbs form}):

\begin{equation} \label{tangent_contact_form}
\boxed{\omega(\vec{t_p}) = \vec{X}.\vec{\dot{x}}} \leq 0
\end{equation}

Where "." denotes the dot product on $\mathbb{R}^n$. \footnote{we will note from now on note the vectors and tensors in bold (in particular $\frac{\partial{f}}{\partial{\vec{x}}}$ denotes the column vector of coefficients $\frac{\partial{f}}{\partial{x_i}} \forall i \in [\![1,n]\!])$} \\

The Gibbs form is thus the scalar product of the affinities vector with the state variables rates vector. As mentioned before, the Gibbs form is required to be non-positive so that the second law can hold, which determines the admissible non-equilibrium processes. The dissipation is defined as the absolute value of the Gibbs form:

\begin{equation} \label{dissipation}
D \equiv -\omega(\vec{t_p})=-\vec{X}.\vec{\dot{x}} \geq 0
\end{equation}

\subsection{Recovering the first and second laws} \label{Recovering the first and second laws}

We can check now that $\omega=0$ and $\omega(\vec{t_p}) \leq 0$ are equivalent to the first and second laws respectively. Given the definition of $\Psi^*$, the Gibbs form can be also written in the form:

\begin{equation} \label{contact_form_bis}
\omega = d\Psi + \vec{y}.d\vec{x}
\end{equation}

For the first law, we aim at retrieving the Pfaffian equation $dU-W-Q=0$. It is written this way to naturally make appear the contact form. When $\omega=0$, this actually corresponds to the differential formulation of the first law, modulo the right contact coordinates. We recommend for instance the didactic explanations in \cite{Salamon2006}.The most obvious choice is to consider the contact coordinates used by Gibbs $(U,P,V,T,S)$ \cite{Gibbs1873} to obtain the (Pffafian form of) the first law $dU+PdV-TdS=0$. Legendre-equivalently, one can use the free energy as the canonical function to get $dF+PdV+SdT=0$. Many other combinations are possible; the different Legendre transforms actually form an infinite dimensional group known as the contact group \cite{Salamon2006}. For instance, $U$ should be used for adiabatic processes whereas $F$ should be used for isothermal processes. The latter will be our choice for the rest of this work. The mechanical deformation term $PdV$ is generalized to $\vec{y}.d\vec{x}$ with $\vec{x}$ the (extensive) deformations of all sorts and $\vec{y}$ the corresponds deforming (intensive) forces. Hence, from now on, we will assume that $\Psi$ is the equilibrium system's free energy, and that $\Psi^*$ is the corresponding generalized free energy.All in all, the first law is embodied by the Legendre submanifolds of the system.\\

As for the second law, the Gibbs form's tangential condition (eq.\ref{tangent_contact_form}) directly yields the dissipation inequality (eq.\ref{isothermal 2 laws}).

When the processes present heat (i.e. non-isothermal), mass or electromagnetic fluxes, the model can be accomodated as in \cite{Haslach2009,Haslach2011}. Indeed, a second contact manifold for the fluxes should be coupled to the existing one, as well as transport one-forms.

\subsection{Maximum dissipation principle} \label{Maximum dissipation principle}

To start with, let us clarify the vector description's wording that we use. We define a vector with three characteristics. Its length is described by its magnitude, its direction corresponds to the angle it makes with a certain reference direction and its sense is where its arrow is pointing at. We apply now the MaxDP on eq.\ref{dissipation} and we require $D$ to be maximum. By the Cauchy-Schwarz inequality, the upper bound of a scalar product is the product of the norms, reached when the two vectors are collinear. Hence, \textbf{the MaxDP is equivalent to requiring $\vec{X}$ and $\vec{\dot{x}}$ to be collinear, of opposite senses.} Let us now define a symmetric definite positive tensor $\mat{\uptau}$ and the associated inner product $<.,.>_{\mat{\uptau}}$:

\begin{equation} \label{inner_product}
<\vec{u},\vec{v}>_{\mat\uptau} \equiv \vec{u}^T \mat\uptau \vec{v} = \vec{u}.(\mat\uptau \vec{v}) \qquad \forall (\vec{u},\vec{v}) \in \mathbb{R}^n\times\mathbb{R}^n
\end{equation}

The associated norm is defined by:

\begin{equation} \label{norm}
||\vec{u}|| \equiv <\vec{u},\vec{u}>_{\mat\uptau}^{1/2}
\end{equation}

The induced inner-product space induces a metric space, whose metric is defined by $d(\vec{u},\vec{v})=||u-v||$. Furthermore, since this inner product space is applied in $\mathbb{R}^n\times\mathbb{R}^n$, it is of finite dimension. The contact manifold $(\mathcal{M},\omega)$ is hence locally a Hilbert space (meaning each point has a neighbourhood homeomorphic to a Hilbert space). More generally, ($\mathcal{M},\omega,\mat\uptau$) is a Riemannian contact manifold. Endowing the contact manifold with a metric $\mat\uptau$ thus allows us to work as in Remannian geometry and to make sense and use of notions such as length, angles or curvatures. This way, we can give a practical meaning to the firstly abstract TPS. Especially, we will be able to speak of a proper distance to equilibrium. 

Working in this new metric space, we constrain $\vec{X}$ and $\vec{\dot{x}}$ to be collinear of opposite direction by taking:

\begin{equation} \label{GRE}
\boxed{\mat\uptau \vec{\dot{x}} \equiv - \vec{X}}
\end{equation}

This tensorial equation corresponds to the $n$ non-equilibrium evolution equations of the system, called relaxation equations if the system is not forced. A system is forced when the control variables $\vec{y}$ are time-dependent, meaning the environment is dynamic, which is the case for systems driven out of equilibrium. Thus we shall call our equation the Generalized Relaxation Equation (GRE). The system "relaxes" from a non-equilibrium state to the long-term (or equilibrium) manifold $\tilde{\mathcal{M}}$ to reach a (dynamic) equilibrium, following the shortest path possible. In reality, we will see that more than reaching a Legendre submanifold, the system can only chase it inasmuch as the submanifold is usually time-dependent, and at best tend towards it. We will show with eq.\ref{GRE_exponential_relaxation} that the still does so as fast as possible.

This formulation leads to a decomposition of the control variables' action into an equilibrium and non-equilibrium (or relaxation) parts:

\begin{equation} \label{control_variable_decomposition}
\vec{y} = -\frac{\partial\Psi}{\partial x} - \mat \tau \vec{\dot x}
\end{equation}

At equilibrium, when the relaxation term vanishes, the usual (equilibrium) Legendre conjugation $\vec{y} = -\frac{\partial\Psi}{\partial \vec{x}}$ is recovered.

Note that the objectivity (frame invariance) of the model is of primal interest, as stressed by Haslach \cite{Haslach2011}, and first by Truesdell \cite{Truesdell1969}. That would allow to model large deformations, which may be very much the case for phases changes. Special care shall then be taken, following the well detailed methodology in \cite{Haslach2011}. Especially, the time rate derivative should be taken as a Lie time derivative if the current configuration does not coincide with the reference configuration. Assuming small deformations as a first approximation in our Lagrangian representation will allow to drop those considerations since the current and reference configurations are considered to coincide. This should still be a mere first approximation.\\

Coming back to our dissipation equation, we check that $D=-\vec{X}.\vec{\dot{x}}=-\vec{\dot{x}}.\vec{X}=\vec{\dot{x}}.(\mat\uptau \vec{\dot{x}})=<\vec{\dot{x}},\vec{\dot{x}}>_{\mat\uptau}$, and thus the dissipation reads simply:

\begin{equation} \label{dissipation_bis}
D = ||\vec{\dot{x}}||_{\mat\uptau}^2
\end{equation}

In terms of Gibbs form, we have:

\begin{equation} \label{gibbs_form}
\omega(\vec{t_p}) = -<\vec{\dot{x}},\vec{\dot{x}}>_{\mat\uptau}
\end{equation}

Thus the two thermodynamic laws qualify $\mat\uptau$ to be a metric of the thermodynamic space (and reciprocally), as noticed in one of the first constructions of a thermodynamic metric \cite{Weinhold1975}. Indeed, $<.,.>_{\mat\uptau}$ defines an inner product if and only if $\mat\uptau$ is symmetric positive definite. The positiveness is ensured by the second law $\omega(\vec{t_p})  \leq 0$ and the definiteness is ensured by the first law since $\omega=0$ corresponds to equilibrium, i.e. $\vec{X}=0$, i.e. $\vec{\dot{x}}=0$. 
Following Haslach's suggestion at the end of \cite{Haslach2011}, we may be tempted to assume that the dissipation measures the distance between the system and equilibrium, i.e. the shortest path or geodesic in terms of differential geometry. Indeed, the dissipation diminishes as the system goes towards equilibrium and vanishes once there: as such the dissipation can be chosen as a Lyapunov function. In other words, we may consider the dissipation as the indicator of a non-equilibrium thermodynamic metric. $\mat\uptau$ totally defines this metric and can be seen as a (generalized) relaxation time tensor, meaning it indicates how long the system takes to go to equilibrium. This temporal metric thus extends the real-world Euclidean measure, in terms of actual distance, to a measure consistent with thermodynamic processes, the thermodynamic time. It can be seen as the distance separating a system to equilibrium but as well as the advancement of its process. \\

To complete the constitutive definition of the model, after defining 1) the system's variables ($\vec{x}$, $\vec{y}$), 2) the equilibrium energy function $\Psi(\vec{x})$, it remains to prescribe 3) the relaxation tensor $\mat\uptau$. It can be defined freely as long as it is symmetric positive definite, and ideally, calibrated with experiments, or least expressed as conventional phenomenological coefficients (e.g. the permeability). It should also be chosen for a satisfying conditioning, i.e. so that $\bar{\theta}_\omega^{max}$ is small enough (cf part \ref{Conditioning of the RMT}). Note that the notion of conditioning introduced theoretically in the present work still remains vague for practical applications.
Note also that this constitutive workflow can be related to the one already preconized by Gurtin \cite{Gurtin1996}, stating that the constitutive relations are fully prescribed when the (equilibrium) energy function and the so-called "kinetic modulus", which we will show equivalent to our relaxation tensor (cf part \ref{Micro-force balance}), are well defined.

\subsection{The relaxation metric tensor}

\subsubsection{How to choose the RMT?}

Note first of all that the dot product, or Euclidean inner product, is recovered in the particular case where $\mat\uptau$ is the identity tensor. Then, a straighforward solution in a first approach is to take $\mat\uptau$ constant and diagonal, assuming a relatively weak coupling of the system's variables. Another solution is to follow the fundamental works in metric thermodynamics like \cite{Weinhold1975} and \cite{Ruppeiner1995}, and take the relaxation tensor as the Hessian of the system's energy, in \cite{Weinhold1975} of the internal energy, in \cite{Ruppeiner1995} of the entropy, or in our case of the free energy. \cite{Ruppeiner1995} took inspiration from the Hessian metric \cite{Weinhold1975} to formally derive a Riemannian geometry for thermodynamics, making the junction with statistical mechanics through a covariant fluctuation thermodynamic theory. In that sense, the RMT can be seen in the light of modern information theory, and even as the thermodynamic limit of the Fisher information metric, which clearly formalizes the connection between such thermodynamic theory and the statistical mechanics at the lower micro-scale. This is to be related to the meaning giving by Haslach \cite{Haslach2011} to the relaxation moduli in his formulation, as containing the information from the lower scale. It should be emphasized however that the previously cited metric theories assume equilibrium mostly, and the metric is used to measure the distance between different equilibrium states. However, here our goal is to measure the distance between a system out of equilibrium and the equilibrium state it is aiming at.

Thus, Riemannian geometry is not formally suited to model non-equilibrium thermodynamics, since it is missing the notion of directionality to follow processes out of equilibrium. While Riemann's geometry quantities are functions of points only of the manifold, Finsler's quantities, seemingly better candidates, are functions of points and directions (since defined on the tangent bundle of the manifold). This confirms the statement of Haslach \cite{Haslach2011} promoting a Finsler metric for non-equilibrium thermodynamics. Note that Riemann's geometry is a particular case of Finsler's geometry.
Valuable insights for the choice of a metric can be found in the so-called geometrothermodynamics framework, launched by Quevedo \cite{Quevedo2006} and followed by Bravetti \cite{Bravetti2013, Bravetti2015}. In particular, a metric should be invariant under Legendre transformations, otherwise the thermodynamic properties of the system depends on the choice of the thermodynamic potential. The previous preliminary insights will not yet be investigated in the present work.

All in all, we have prescribed qualitatively the relaxation path using the MaxDP but not yet quantitatively since the relaxation tensor, i.e. the metric is not explicitly defined. It is surely not an easy task as the metric carries a deep physical and mathematical meaning which has not clearly be settled yet, albeit numerous attempts.


\subsubsection{Conditioning of the RMT} \label{Conditioning of the RMT}

We will provide nonetheless in this part a new insight regarding the RMT using our framework. It is convenient and visual to follow the thermodynamic process via the Gibbs contact angle $\theta_\omega$. Recall that $\theta_\omega$ is the obtuse angle between $\vec {\dot{x}}$ and $\vec{X}$. For convenience, we can consider instead the complementary Gibbs angle $\bar{\theta}_\omega=\pi-\theta_\omega$, the acute angle between $\vec {\dot{x}}$ and $-\vec{X}$ (see fig.\ref{fig:conditioning_bounding}). The MaxDP amounts to vanishing $\bar{\theta}_\omega$. Using the conditioning bound lemma derived in \ref{conditioning bound derivation}, since $\mat\uptau$ is symmetric definite positive, we get the following upper bound for the complementary Gibbs angle:

\begin{equation} \label{conditioning_bounding}
\bar{\theta}_\omega \leq cos^{-1}(\frac{\lambda_m}{\lambda_M})=cos^{-1}(cond(\mat\uptau)^{-1})=\bar{\theta}_\omega^{max} \in [0,90 ^\circ]
\end{equation}

Where $\lambda_m$ and $\lambda_M$ are respectively the minimum and maximum eigenvalues of $\mat\uptau$ (both strictly positive since $\mat\uptau$ is definite positive). The conditioning bounding can be visualized below in fig.\ref{fig:conditioning_bounding}:

\begin{figure}[h!]
\centering
\includegraphics[scale=0.2]{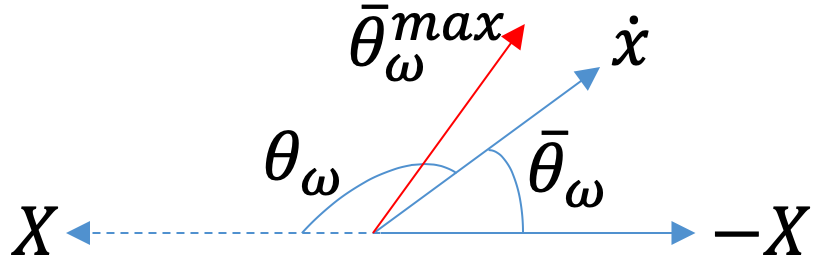}
\caption{Conditioning bounding}
\label{fig:conditioning_bounding}
\end{figure}

The condition number is usually used in numerical analysis and defines the sensitivity of a function's output to a variation of the input. If this number is large, the problem is said ill-conditioned and if it is close to 1, well-conditioned. By analogy, the thermodynamic condition number can be interpreted as quantifying the sensitivity of the thermodynamic path to fluctuations. Indeed, if it is large, the Gibbs angle has a large range of variations. Conversely, if the condition number is close to 1, the Gibbs angle is constrained to small values. Hence, it seems that far from equilibrium, a process should be ill-conditioned, subject to fluctuations and deviation from the geodesic, whereas close to equilibrium, it should be well-conditioned. In the first case, the system tends to a chaotic regime, in the second case, to a deterministic regime. Physically, given the definition of $\mat\uptau$, the first case is due to very different relaxation times for the respective system's variables, whereas for the second case, the different variables have similar relaxation times. The link between reaction rates discrepancy and patterns formation (happening out of equilibrium) has already been highlighted in reaction-diffusion systems' theory (cf part \ref{Reaction-diffusion systems}). 

A last parallel can be drawn between our conditioning bound and information theory. From our thermodynamic point of view, the information that the system uses to progress are thermodynamic forces, i.e. the affinities vector $\vec X$, indicating the closest equilibrium manifold at any time (the direction is given by $-\vec X$). When $\bar{\theta}_\omega$ is large, the system gets away from the information line (the affinities vector) and it becomes more unpredictable. Conversely, if $\bar{\theta}_\omega \approx 0$ the system goes in line with the provided information and is fully predictable. Recall that we understand predictable for a system when it is close enough to an equilibrium manifold (i.e. a steady state); that is when the system is at rest (and observable/measurable), with the proviso that the steady or equilibrium state is stable enough to allow observation. 

All in all, this insight on conditioning bounding shows that the GRE, in appearance a trivial linear relationship, allows deviations from the ideal path provided by the MaxDP.

\subsection{Generalization of Onsager's relations?}

In this part and the next two parts, we draw comparisons and intent to shed some light on some major paradigms in thermodynamics, namely Onsager's, Prigogine's and Bejan's contributions.

To start with, let us derive a Onsager-like formulation \cite{Onsager1931}. For that, we can write the fluxes $\vec{J}$, i.e. the rate of the state variables, and recover:

\begin{equation} \label{onsager_fluxes}
\vec{J} \equiv \vec{\dot{x}} = - \mat\uptau^{-1} \vec{X} = \mat{L} \vec{X}
\end{equation}

At first sight, this Onsager-like equation derived from the GRE seems similar to the original one, in the sense that it yields a linear relationship between the affinity $\vec{X}$ and the flux $\vec{J}$. Notwithstanding, X major (related) differences arise. First, we consider generalized affinities $\vec{X}=\frac{\partial \Psi^*}{\partial \vec{x}}$ (via the generalized free energy) rather than equilibrium affinities $\tilde{\vec{X}}=\frac{\partial \Psi}{\partial \vec{x}}$ (via the equilibrium free energy), with $\vec{X}=\vec{\tilde{X}}+\vec{y}$. Note however that Onsager used the entropy as system's energy. The difference is thus that we take into account the dynamic influence of the environment (provided that $\vec{y}$ is time-dependent). Second, the linearity comes from the MaxDP rather than assuming near-equilibrium. In relation with the previous point, this near-equilibrium hypothesis is retrieved in the use of $\tilde{\vec{X}}$, which described the thermodynamic forces on (or near) the Legendre submanifolds. Third, the symmetry of the phenomenological coefficient $\mat{L}$ in our framework stems from the metric structure, support for thermodynamic irreversibility (cf part \ref{source_of_irreversibility}), whereas the symmetry of Onsager's coefficient comes from the assumption of microscopic reversibility \cite{Onsager1931}.

\subsection{Recovering the source of irreversibility beyond Hilbert's space} \label{source_of_irreversibility}

We now intent to explain the source of irreversibility from the contact structure $ker(\omega)$ in the light of Prigogine's ideas. We remind that the contact structure results from an abstract mathematical construction aiming at embedding the two first laws of thermodynamics into a maximally non-integrable structure living in the tangential space of the system. The maximal non-integrability condition provides the contact structure with a fundamentally discrete structure. Essentially, the information is transmitted from the contact structure to the manifold by an ideal tangential approximation in the sense that the MaxDP indicates to the process the shortest path possible in between two consecutive states. But that comes with a price.

As already famously formulated by Progogine, "to grasp the real world, we must leave Hilbert space" and complete it with "a holistic, nonlocal description" so that "irreversibility is incorporated into the laws of nature" \cite{Prigogine1996}. In our case, the Hilbert physical space is associated with the (Riemannian contact) TPS - the latter is locally Hilbertian. The irreversibility, or "arrow of time", appears in the Lyapunov structure of the GRE (the function $\mathcal{D}=-\omega(\vec{t_p}) \geq 0$ decreasing until vanishing on Legendre submanifolds  can be considered as a Lyapunov function). This feature is directly related to the relaxation metric $\mat\uptau$, geometrical representation of the second law. We here part ways with the linear representation of the arrow of time, based on the assumption that entropy perpetually increases. As discussed in part \ref{Conditioning of the RMT}, the latter is only valid for the very limiting cases of equilibrium states, i.e. for statics, merely pointing out that a system will die if left alone. Temporal evolution, i.e. dynamics, deals with non-equilibrium processes.

More fundamentally, in terms of information theory, one could say that the information, discrete by nature, lives in the contact structure, the latter thus encoding the physical Hilbert space, via the TPS. Recall that the Hilbert space and the TPS are homeomorphic (the latter is locally the former). The smoothing of the discrete (maximally non-integrable) contact structure into the continuous TPS thus comes with the price of information loss. As a result, the system would not be able to recover its path back in time. Now relating to the problematic raised by Prigogine, irreversibility can be seen as incorporated into the contact structure, inherently non-local and beyond the Hilbert space. In that sense, time would be all but a smooth linear arrow-like process. Its perceived asymmetry would stem from the dichotomy between the virtual information that has been actualized into processes (which cost is irreversibility) and that has not been. 

All in all, we can distinguish three levels, the contact structure (world of information), the TPS (world of processes) and the Hilbert space (world of observations). The first encodes the second, which is observable from the third (cf fig.\ref{fig:sketch_contact_thermodynamics}).

%

\subsection{Break/engine behavior, analogy with the constructal law} \label{Break/Engine behavior}

Our contact thermodynamic framework can be summarized with the dissipation equation:

\begin{equation}
\mathcal{D} = \mathcal{P} + \Sigma \equiv D_{max} = ||\dot{\vec{x}}||_{\mat\uptau}^2
\end{equation}

With $\mathcal{D}=\vec{X}.\dot{\vec{x}}$ (dissipation), $\mathcal{P}=-\dot\Psi$ (self-organization power), $\Sigma=-\vec{y}.\dot{\vec{x}}$ (entropy production).\\

A process is the result of the interplay between the thermodynamic velocity $\dot{\vec{x}}$ and thermodynamic forces: the former is required to align with the latter at all times, knowing that the latter is moving if the equilibrium manifolds are moving, i.e. if the environment is soliciting the system (if the control variables are time-dependent). If at an instant $t$ the two vectors are aligned, but at the next instant $t+dt$, $\vec{X}$ changes direction, then the system will have to spend an amount of energy $\mathcal{D}$ to realign with the optimal path to equilibrium (cf fig.\ref{fig:sketch_contact_thermodynamics}). The system can do so in two (extreme) ways:\\

1) \textbf{Brake behavior (MaxEPP)}: 

If $\Sigma \approx \mathcal{D}$ (i.e. if the system is already fitted to fully process the environment energy input), the energy input is fully dissipated back to the environment. There is no need for the system to self-organize ($\dot\Psi \approx 0$). Thus the dissipation is maximized by maximizing the entropy production.\\

2) \textbf{Engine behavior (MinEPP)}:

Conversely, if the system is not fitted at all to process the energy input ($\Sigma \approx 0$), the system will have to completely readapt its organization ($\dot\Psi \approx -\mathcal{D} \leq 0$). This behavior can be related to Prigogin's dissipative structures following MinEPP. Indeed, when most of the input energy is used for self-organization, the entropy production $\Sigma$ can be seen as kept to a minimum. In other words, the dissipation is maximized by minimizing the entropy production (or maximizing $\mathcal{P}$). \\

Note that we took direct inspiration for the denomination from Bejan's constructal theory \cite{Bejan2016}, pushing further the parallel between our models.
Obviously an actual behavior is a mix of the two previous ideal cases. A third characterisitic behavior can be the storage behavior, extension of the engine behavior. Then $\dot\Psi \geq 0$ and the system stores some useful energy than it can use for future self-organization. This can be assimilated to the so-called "cold work". However this is possible on the proviso that $\dot\Psi \leq \Sigma$. This third behavior could be interpreted as the system being overfitted ($\Sigma \geq \mathcal{D}$) and able to stored the excess energy ($\Sigma - \mathcal{D}$) as cold work ($\dot\Psi$).   \\

Now let this insight shed light on the confusion around extremal principles mentioned in the introduction. As stated in \cite{Martyushev2013}, the MaxEPP prevails on a short time scale whereas the MinEPP on the longer time scale. This makes sense intuitively since dissipating straight away the energy input is faster for the system than having to reorganize its way of processing it before dissipating it. Thus assuming a clear separation of those two time scales, the MaxDP reduces to the MaxEPP ($\mathcal{D} \approx \Sigma$) on the short time scale and to the MinEPP ($\mathcal{D} \approx \mathcal{P}$) on the long time scale. Further, let us consider the long term manifold characterized by this long time scale. We can show that our GRE can be written in the form: 

\begin{equation} \label{GRE_affinity_form}
\dot{\vec{X}}=-\mat{R} \vec{X}
\end{equation}

With $\mat{R}=\mat{H}\mat\uptau^{-1}$ and $\mat{H}=\frac{\partial^2{\Psi^*}}{\partial \vec{x}^2}=\frac{\partial^2{\Psi}}{\partial \vec{x}^2}$ (Hessian).\\

As we are considering the long term manifold, assuming the system is tending to a stable equilibrium state, we can consider $\mat{R}$ to be constant in time (at least close enough to the long term manifold) and definite positive. Therefore the solution is:

\begin{equation} \label{GRE_exponential_relaxation}
\vec{X}(t) = e^{-t \mat{R}} \vec{X_0}
\end{equation}

Where we use the usual exponential of matrices and $\vec{X_0}$ is the affinities vector at the assumed starting time of convergence to the long term manifold (i.e. from when $\mat{R}$ can be considered constant).\\

\begin{figure}[h!]
\centering
\includegraphics[scale=0.3]{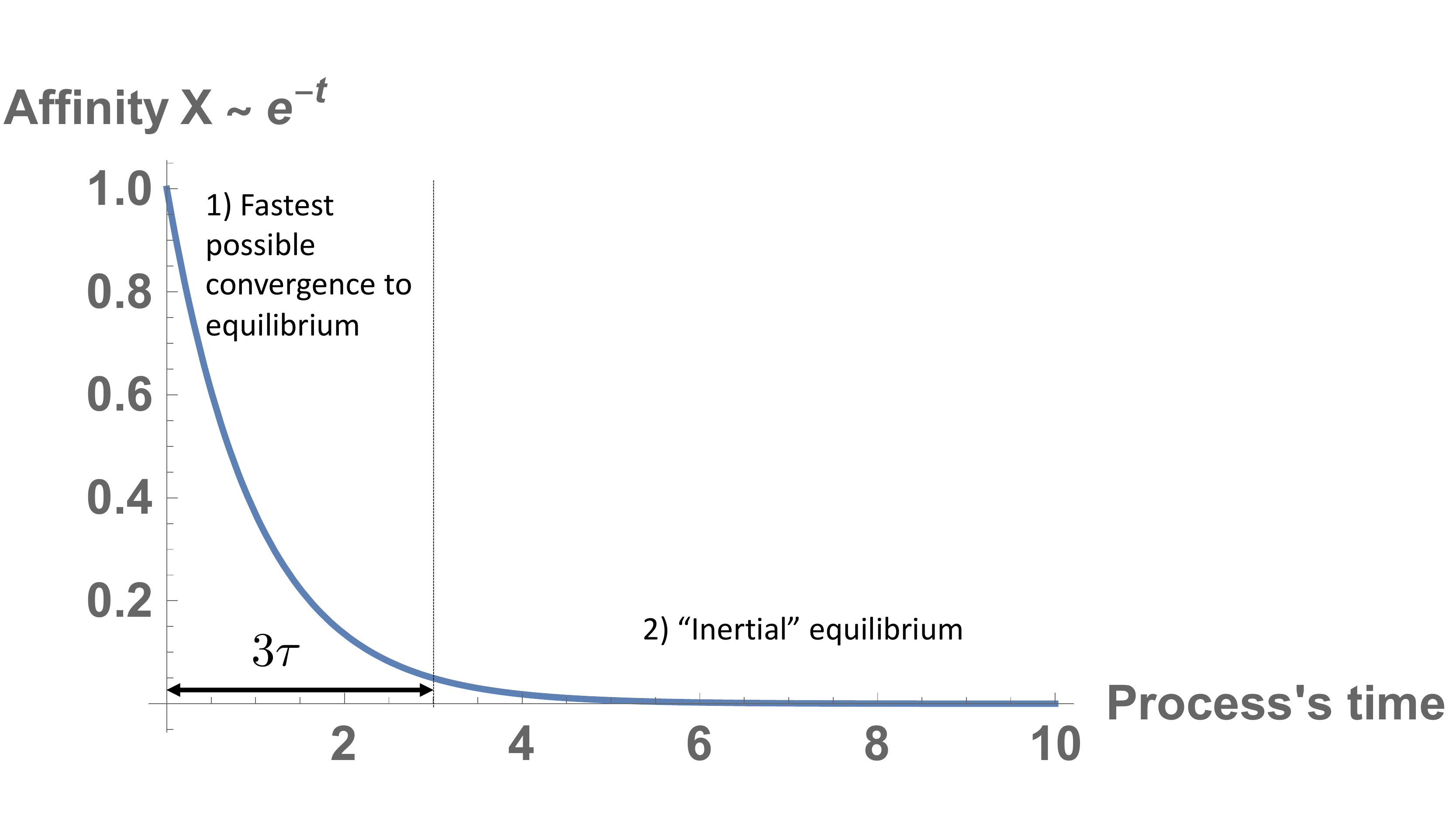}
\caption{Exponential convergence to equilibrium in the simple case $X(t)=e^{-t}$. A first regime of duration $3\tau$ e.g. ($\tau=1$ here) corresponds to the fastest convergence possible to equilibrium. An "inertial" equilibrium can be approximated by a second regime where the system tends to $0$.}
\label{fig:convergence_eqbm}
\end{figure}

Four points can be straightforwardly made on GRE2. First, the MaxDP implies an exponential convergence to equilibrium (if stable), which the system shall reach at infinite time. This formally proves our claim that the MaxDP pushes the system to reach equilbrium as fast as possible. Second, GRE2 corroborates the fact that the stability of the Legendre (equilibrium) submanifolds is determined by the nature of the Hessian $H$. Third, the relaxation tensor $\mat\uptau$ (which coefficients are relaxation times associated to the different state variables) takes all its sense since, although in tensorial form, it can be seen as dividing the time of the exponential decay. Fourth, this could also corroborate Prigogine's idea of "kind of 'inertial' property of nonequilibrium system" \cite{Prigogine1978} to settle at a state of least entropy production without being able to reach full equilibrium. Recall that $\vec{X}$ is directly related to $\mathcal{D}$ and thus the latter reaches zero iff the former does ($\mathcal{D}=||\vec{X}||^2_{\mat\uptau^{-1}}$). We can summarize those observations simply on the graph on the decreasing exponential function $e^{-t}$ (cf fig.\ref{fig:convergence_eqbm}).

Furthermore, this engine/brake illustration pushes further the analogy with the constructal law. Indeed, Bejan describes the constructal law as "the natural tendency of evolution toward flow configurations that provide easier access to what flows." The idea that a system should "go with the flow" can seemingly be formalized with CT, where the system has to "go with the flow", in the sense that the thermodynamic path ($\vec{\dot{x}}$) corresponds to the thermodynamically preferred path ($\vec{X}$), modulo the relaxation metric $\mat\uptau$. The correspond collinearity of $\vec{\dot{x}}$ and $\vec{X}$ happens to correspond to the maximization of the dissipation.\\

\subsection{Summary of contact thermodynamics}

To close this part on the construction of CT, we propose a sketch here below describing a contact thermodynamic process in "1D", the simplest case for visualization (i.e only one thermodynamic couple $(x,y)$). Recall that the TPS is a $2n$-dimensional submanifold (locally the graph of $\Psi^*(\vec{x},\vec{y})$) of the $2n+1$-dimensional contact manifold $(\mathcal{M},\omega)$ (described the coordinates $(\Psi^*,\vec{x},\vec{y})$). The 2n-dimensional contact structure $\ker(\omega)$, a field of hyperplanes in the tangent space, "maximally tangents" (since $\omega$ is maximally non-integrable) the TPS. This contact structure ensures the "contact" between the system and its environment and embodies the first law ($\omega=0$) and the second law $\omega(\vec{t_p}) \leq 0$. In this contact structure lie the Legendre (or equilibrium) submanifolds of dimension $n$, maximal solution of $\omega=0$ (first law). The information from the discrete contact structure is translated into continuous thermodynamic processes in the TPS via the MaxDP. Thus, the system maximally tends (cf eq.\ref{GRE_exponential_relaxation}) to the closest Legendre submanifold to reach (or rather tend to) equilibrium. 

\begin{figure}[h!]
\centering
\includegraphics[scale=0.35]{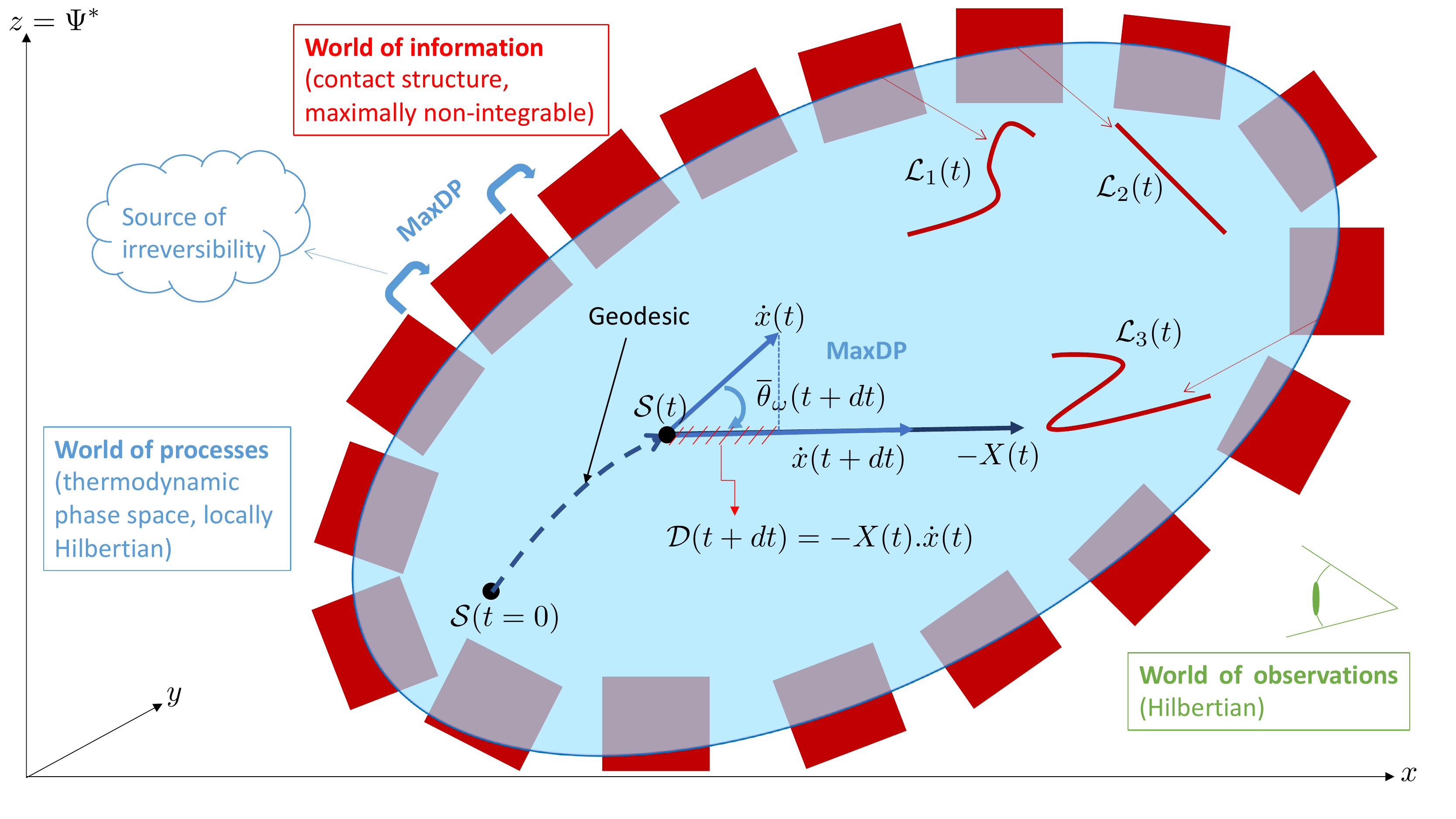}
\caption{Representation of the 2D-TPS in a 3D-contact manifold, along with its associated 2D-contact structure. The discrete contact structure, which can be seen as encapsulating the information, is smoothed out via the MaxDP into the TPS containing the continuous thermodynamic paths. At each instant, the systems heads ($\dot{x}(t)$) for the closest way to equilibrium, i.e. for the closest Legendre submanifold, which direction is given by the affinity $X(t+dt)$. The resulting path is called a geodesic. The price to pay for the system to readjust to the optimal path to equilibrium is the dissipation (source of irreversibility). The TPS, locally Hilbertian, is related to the Hilbertian physical space (via homeomorphism).}
\label{fig:sketch_contact_thermodynamics}
\end{figure}

Hence, in the simplest 1D case ($n=1$), the TPS can be represented in the 3D space charted with the contact coordinates $(\Psi^*,x,y)$ by a surface (2D). The field of 2D hyperplanes tangenting the TPS represents the contact structure. Therein can be found the 1D Legendre submanifolds (lines), projected on the TPS for better legibility. We represent here below (\ref{fig:sketch_contact_thermodynamics}) an attempt of making clearer graphically most of the concepts of CT. It was to our knowledge missing in the literature and further clarifications may be required.

(cf fig.\ref{fig:sketch_contact_thermodynamics}).

That draws a conclusion to the construction of our CT framework. As mentioned earlier, thermodynamics is a black-box theory. As explained by Salamon in \cite{Salamon2006}, contact geometry clearly formalized the intuition that a system can be (partially) understood by choosing a certain set of control/environmental variables ($\vec{y}$) and see how the system reacts to it by observing the corresponding state variables ($\vec{x}$). The contact form ($\omega$) provides the adequate tool to make this connection. To further understand internal processes like self-organization and go beyond the black box, theories like PFM modeling internal phase changes can be helpful. This is what we will attend to in the next part.

\section{Application to phase-field modeling}

If one agrees that PFM deals with non-equilibrium processes, it is legitimate to check whether its conventional framework corroborates it. After discussing such a fundamental issue in the introduction, we may now derive a PFM fully embedded in a non-equilibrium thermodynamic framework, and see what may change. For that, we opt for the previously introduced CT. The association of PFM with CT is not trivial. Both can be seen complementary since the CT describe the contact or interaction of a black box system with its environment, whereas PFM describes the internal structure of a system. We thus aim at overcoming the phenomenological nature of thermodynamics.

\subsection{State and control variables}

First we have to choose the state and control variables. Obviously we have to choose $x_1 \equiv \phi$ where $\phi$ is a general order parameter, differentiating different phases. Now, the essence of PFM is that it is a higher-order (gradient) theory with respect to the sharp interface model. The introduction of the gradient of the order parameter in the system's free energy allows to deal with diffuse interface, rather than discontinuous interfaces. Hence by definition, $\nabla\phi$ should be taken as an independent state variable, in addition to $\phi$, i.e. $x_2 \equiv \nabla\phi$. We associate the respective control variables $y_1$ and $y_2$, yet to be specified.

\subsection{Equilibrium free energy}

Following \cite{Cahn1958}, we Taylor-develop the system's free energy up to second order to make appear $\nabla\phi$. Assuming isotropy, the odd orders and off-diagonal terms vanish and we get:

\begin{equation} \label{free_energy_basic}
\Psi(\phi,\nabla\phi) = B(\phi) + \frac{\Gamma}{2}||\nabla\phi||^2
\end{equation}


Where $B(\phi)$ is the bulk energy (in $J/m^3$), or the energy as if the material were homogeneous, $\Gamma=\gamma l_i$ is the interfacial energy (in $J/m$), $\gamma$ is the surface tension (in $J/m^2$) and $l_i$ a characteristic interface thickness. The gradient term penalizes the order parameter variations, and in fact penalizes the artificial introduction of interface thickness. In other words, a highly fluctuating order parameter is energetically less favorable than an order parameter with little variations. Another characteristic of PFM is that the bulk energy term should contain a double-well potential ensuring that the system tend to stabilize by choosing between the different phases and not a state in between (see the linear stability analysis in part \ref{LSA}). To destabilize the system a source energy term should appear in the bulk energy as well. Thus:

\begin{equation} \label{bulk_energy}
B(\phi) = Gg(\phi) + \overline{H}(\phi)
\end{equation}

$Gg(\phi)$ is the classical double-well potential with $G$ the height of the double well and $g(\phi)=\phi^2(1-\phi)^2$, assuming the two stable phases correspond to $\phi=0$ (phase A) and $\phi=1$ (phase B).  $\overline{H}(\phi)$ is the mixture of the bulk energy of the different phases, here A and B. For that, we use an interpolation function $h(\phi)=\phi^2(3-2\phi)$. Thus the free energy reads:

\begin{equation} \label{free_energy_interpolation_function}
\Psi(\phi,\nabla\phi) = Gg(\phi) +(1-h(\phi))H_A + h(\phi)   H_B + \frac{\Gamma}{2}||\nabla\phi||^2
\end{equation}

Note that for now the bulk source energies $H_A$ and $H_B$ are supposed constant.

\subsection{Micro-force balance} \label{Micro-force balance}

We follow the fundamental microforces theory developed by Fried and Gurtin, initially introduced in \cite{Fried1993}, and detailed in \cite{Gurtin1996}.

While the CT framework based on the MaxDP provides evolution laws relating the state variables ($\phi$, $\nabla\phi$) and the respective control variables ($y_1$, $y_2$), one equation for each thermodynamic pair (so two in the case of PFM), one would rather work with a more practical equation containing only $\phi$ and $\nabla\phi$. For that, a fundamental law relating $y_1$ and $y_2$ shall be supplemented. We naturally choose the "balance of accretive forces" \cite{Fried1993}, the term accretion meaning "the growth of one phase at the expense of another" \cite{Fried1994}. It is called as well "microforces balance" \cite{Gurtin1996}, phrase that we adopt. Fried and Gurtin's theory stems from the assumption that "fundamental physical laws involving energy should account for the working (or power expenditure) associated with each operative kinematical process" \cite{Gurtin1996}. It is important to note that generally only the kinematics associated with the order parameter ($\dot\phi$) is considered (and not its gradient $\nabla\dot{\phi}$). Their "microforce system" is described by a scalar microforce $\pi$, power-conjugated to $\phi$ and a vector microstress $\vec{\xi}$, power-conjugated to $\nabla\phi$. Writing the expenditure of power within an arbitrary control volume $\mathcal{R}$, with $\vec{n}$ the outward unit normal to $\partial \mathcal{R}$ gives the so-called microforce balance (neglecting the external sources) in non-local and local forms:

\begin{equation} \label{non_local_microforce_balance}
\int_{\partial\mathcal{R}} {\mat{\xi}.\vec{n} dS} + \int_{\mathcal{R}} {\pi dV} = 0
\end{equation}

\begin{equation} \label{local_microforce_balance}
\nabla.\xi+\pi = 0
\end{equation}

It is fundamental to discuss the choice of the dissipative kinematics. As per Fried and Gurtin's derivations, the classical Allen-Cahn and Cahn-Hilliard equations assume implicitly that only the microforce associated with the order parameter variations is dissipative. However, by definition, the PFM adds $\nabla\phi$ to the system's energy, independently from $\phi$. Hence the working associated with the kinematics of $\nabla\phi$ should be taken into account as well. It is done by Gurtin in \cite{Gurtin1996} by considering a dissipative, or viscous, microstress, that is, in their formulation, by including $\nabla\dot\phi$ in the constitutive variables. Yet, it seems that the dissipation related to $\nabla\phi$ should be more than a possible add-on to the Allen-Cahn equation. Indeed, if the working of each kinematical process should be taken into account, it appears that a PFM should consider a dissipative microstress in addition to a dissipative microforce. Even from a more basic point of view, in order to model interfacial problems, the system should be allowed to move and dissipate through two dimensions, the interface normal variations ($\dot\phi$) and the interface orientation variations ($\nabla\dot{\phi}$). Nonetheless, it may be justified to neglect the rate term related to curvature variations (for processes with low variations of interface curvature), but it should at least be considered.

In \cite{Gurtin1996}, the dissipation equation, after applying the Coleman-Noll procedure, assuming a viscous stress (i.e. including $\nabla\dot\phi$ as a constitutive variable), is written in the form:

\begin{equation}
\vec{F}(\vec{X},\vec{Y}).\vec{Y} \leq 0
\end{equation}

That we may rewrite for analogy to our model:

\begin{equation}
\vec{F}(\vec{x},\vec{\dot{x}}).\vec{\dot{x}} \leq 0
\end{equation}

Where $\vec{x}=(\phi,\nabla\phi)$, $\vec{F}(\vec{x},\vec{\dot{x}})=(\pi_d,-\vec{\xi_d})$, $\pi_d=\pi+\frac{\partial\Psi}{\partial\phi}$, $\vec{\xi_d}=\vec{\xi}-\frac{\partial\Psi}{\partial\nabla\phi}$.\\

By taking $y_1=\pi$ and $\vec{y_2}=-\vec{\xi}$, $\pi_d$ and $\vec{\xi_d}$ are actually what we called the affinities $X_1$ and $\vec{X_2}$, and then we recover our Gibbs form inequality:

\begin{equation}
\vec{X}.\vec{\dot{x}} \leq 0
\end{equation}

Gurtin shows in \cite{Gurtin1996} that then there exists a matrix $\mat{B}(\vec{x},\vec{\dot{x}})$ with $\mat{B}(x,0)$ positive semidefinite, such that $\vec{F}(X,Y)=-\mat{B}(X,Y).\vec{Y}$, i.e. $\vec{X} = -\mat{B}(\vec{x},\vec{\dot{x}})\vec{\dot{x}}$. Recall that we establish that collinearity via Cauchy-Schwarz's argument ($\mat\uptau\vec{\dot{x}}=-\vec{X}$). Thus Gurtin's kinetic tensor $\mat{B}$ corresponds to our relaxation tensor $\mat\uptau$. We may therefore think that the constitutive kinetic assumption of Gurtin corresponds actually to the MaxDP. However, Gurtin obtained a weaker assumption than positive definiteness for $\mat{B}$, reduced to positive definiteness when $\mat{B}$ is independent from the rates of the variables. Note that positive definiteness is not compulsory either in our model but is enforced to get a metric and make sense of it in our geometrical framework.\\



\subsection{CPFM with constant bulk energy}

To start with and emphasize the consistent structure of our CPFM, especially the separation of the fundamental laws (thermodynamics and balances) and the constitutive assumptions, we may use the following table. It will be useful as well to keep things clear when extending the model to supplementary physics.\\

\begin{table}[!htbp]
\centering
\caption{CPFM with constant bulk energy}
\label{CPFM with constant bulk energy}
\begin{tabular}{|c|c|c|c|}
\hline
\multicolumn{2}{|c|}{Thermodynamic Framework}                              & Supplementary Balance Laws & Constitutive Assumptions \\ \hline
State Variables                      & Control Variables                     & Micro-force     & $\Psi(\vec{x})$              \\ \cline{1-2}
$\phi$                               & $\pi$                                 &          & MaxDP ($\mat\uptau \vec{\dot{x}} = - \vec{X}$)                      \\
$\nabla\phi$                         & $-\xi$                                &                            & $\mat\uptau$ (metric)                               \\ \cline{1-2}
\multicolumn{2}{|c|}{1st law: $\omega=0$ on equilibrium manifolds} &                            &                          \\ \cline{1-2}
\multicolumn{2}{|c|}{2nd law: $\omega(\vec{t_p}) \leq 0 \iff$ CDI}           &                            &                          \\ \hline
\end{tabular}
\end{table}

Applying eq.\ref{GRE}, the two GREs for PFM read:

\begin{equation}
\tau_1 \dot{\phi} = - (Gg'(\phi)+[H_B-H_A]h'(\phi) + \pi)
\end{equation}

\begin{equation}
\tau_2 \nabla\dot{\phi} = - (\Gamma\nabla\phi - \mat\xi)
\end{equation}

To collapse $\pi$ and $\mat\xi$, we use the micro-force balance $\nabla.\mat\xi + \pi = 0$ introduced by Fried and Gurtin \cite{Fried1993} and we get the CPFM:

\begin{equation}
\boxed{-\tau_2\Delta\dot{\phi} + \tau_1\dot{\phi} = \Gamma\ \Delta\phi - f(\phi)}
\end{equation}

Where $f(\phi) = B'(\phi) = Gg'(\phi)+[H_B-H_A]h'(\phi)$ is the bulk energy term.\\

This equation is similar to that obtained by Gurtin \cite{Gurtin1996} when considering a viscous microstress, i.e. adding $\nabla\dot\phi$ in the list of constitutive variables. As previously discussed, we can infer that Gurtin's and Fried's model, using the Coleman-Noll procedure, is not a priori restricted to close-to-equilibrium processes and implicitly assumes the MaxDP.

\subsection{Comparison with the variational formulation} \label{comparison with variational formulation}

It is interesting to now pursue the comparison between our formulation and the variational formulation or gradient flow equation (cf eq.\ref{variational_PFM}), since both models have a priori similar assumptions, especially relaxation to equilibrium, but our formulation ends up with an additional term, the Laplacian rate $\Delta\dot\phi$. The variational formulation uses the free energy as a Lyapunov function to minimize it. That makes sense inasmuch as this corresponds to a perpetual self-organization and PFM is interested in phase changes. However our free energy has no reason to be necessarily decreasing ($\dot\Psi = \Sigma -\mathcal{D}$ with $\Sigma \geq 0$ and $\mathcal{D} \geq 0$). Our requisite is that $\mathcal{D}$ be maximized, which can be achieved in various ways not necessarily by minimizing the free energy, which we call the brake behavior (cf discussion in part 2.8.). We can still note that our generalized free energy $\Psi^*$ could be taken as a Lyapunov function and minimized when there is no change in energy input from the environment (i.e. $\vec{y}$ is constant in time: $\dot\Psi^* = -\mathcal{D}-\vec{x}.\vec{\dot{y}}=-\mathcal{D} \leq 0$ (with $\mathcal{D}$ maximized).

Despite those significant differences in founding assumptions, the final PFM equations are still similar, with the only difference being the Laplacian rate term. We indeed formally recover the classical relaxation equation, with $\tau_2=0$:

\begin{equation}
-\tau_2\Delta\dot{\phi} + \tau_1\dot{\phi} = -\frac{\partial\Psi}{\partial\phi} +\nabla.\frac{\partial\Psi}{\partial\nabla\phi}=-\frac{\delta\Psi}{\delta\phi}
\end{equation}

To close the comparison with the variational formulation, we insist that our formulation that follows the microforces formulation of Fried and Gurtin grants two major advantages already detailed in their work \cite{Fried1993}. Firstly, it gives more freedom in modeling especially in the way the rate terms should be included or not in the basic equations. The variational formulation does not allow in particular the dynamics of the gradient of the order parameter, although by essence of PFM it is a full-fledged state variable. Gurtin associated it to considering a viscous micro-stress \cite{Gurtin1996}. Secondly, our formulation clearly separates the model's constitutivity from the balance laws. We emphasize it in the CPFM tables formulation, separating the thermodynamic framework from the supplementary balance laws and from the constitutive assumptions. We also hope to have provided a physical meaning to the constitutive assumption made in Fried and Gurtin's work, namely the MaxDP.

\subsection{Mechanical coupling}

We now couple the PFM with mechanics by taking the bulk source energy as the mechanical energy and including the strain tensor in the state variables, coupled with (Cauchy) stress tensor as control variable. Major assumptions are made to keep things simple as a first step, but we shall remain aware of them and look into waving them in a second step. 1) No plasticity is considered yet. Thus $\vec{\epsilon}=\vec{\epsilon}^e$ will implicitly denote the elastic strain We may consider that the microstructural scale we will be working at (geomaterials grain scale are considered in the second part of this work \cite{Guevel2019b}) describes the micro-physics that plasticity averages and therefore we may consider plasticity only in the upscaling of the PFM. Indeed the irreversibility is already included in the normal and orientation changes of the phases interfaces, whereas the mechanics holds only in the phases bulk and thus dissipates through the interfaces movements. Even though plasticity was to be considered at the grain scale, we assume that elasticity would remain the driving force and plasticity only an energetic sink term that would only delay the process. 2) We assume the macro-force (or macro-momentum) balance to hold, since the diffusion of the mechanical energy is assumed much faster than the phase field diffusion - the mechanics can accommodate at each time the change in phase field but not the opposite. Thus we neglect the macroscopic inertial effects but we are aware that it may be important to include them later. 3) We assume small deformations to get the current configuration to coincide with the reference configuration and ensure frame invariance (see discussion on the time derivative in part \ref{Micro-force balance}). It is however unlikely that the deformations remain small enough during phases changes, so the model should consider including later finite deformations or move on to a Eulerian formulation. 4) We consider the phase A to be the mechanically weak phase and phase B the strong phase. As an application in \cite{Guevel2019b}, we will consider the phase A as the pore phase in a geomaterial and the phase B as the matrix/grains. The pores fluid (air and/or liquid) will be then taken as a shear-free solid much more deformable than the matrix phase. It is a first approximation and ideally the mechanics should be coupled with hydrodynamics. This solid representation of a liquid or gas phase is made as well in \cite{Kassner2001} for instance. 5) Finally, each phase is considered to be a homogeneous isotropic material with their own free energy.\\

For each phase $K$, the mechanical energy becomes:

\begin{equation}
H_K=\frac{1}{2}\mat{\epsilon}.\mat{C}\mat{\epsilon} = \frac{1}{2}C_{ijkl}^K\epsilon_{ij}^{K}\epsilon_{kl}^{K}
\end{equation}

With $C_{ijkl}^K=\lambda^K\delta_{ij}\delta_{kl}+\mu^K(\delta_{ik}\delta_{jl}+\delta_{il}\delta_{jk})$ the elastic tensor, $\lambda$ and $\mu$ being the Lam\'e parameter and the shear modulus respectively. 

%

The equilibrium free energy is the same except that the strain is now to be counted among the state variables:

\begin{equation}
\Psi(\phi,\nabla\phi,\vec{\epsilon}) = Gg(\phi) +H_A(\vec{\epsilon})(1-h(\phi))+  H_B(\vec{\epsilon})h(\phi) + \frac{\Gamma}{2}||\nabla\phi||^2
\end{equation}

We choose a Voigt homogenization scheme (see e.g. \cite{Ammar2009}), i.e. we assume homogeneous strains: $\vec{\epsilon_A}=\vec{\epsilon_B}=\vec{\epsilon}$. Then the stress of each phase is computed following Hooke's law $\mat{\sigma_K}=\mat{C}_K\mat{\epsilon}$ and the homogenized stress is, in theory, interpolated as per Voigt's scheme: $\mat\sigma(\phi) = h(\phi)\mat{\sigma_B} + (1-h(\phi))\mat{\sigma_A}$. However, we will see that our model yields an additional viscous term to the homogenized stress expression. \\

The CPFM is now:

\begin{table}[!htbp]
\centering
\caption{CPFM with mechanical coupling}
\label{CPFM with mechanical coupling}
\begin{tabular}{|c|c|c|c|}
\hline
\multicolumn{2}{|c|}{Thermodynamic Framework}                              & Supplementary Balance Laws & Constitutive Assumptions \\ \hline
State Variables                      & Control Variables                     & Micro-force Balance     & $\Psi(x_i)$              \\ \cline{1-2}
$\phi$                               & $\pi$                                 &    Macro-force Balance      & MaxDP ($\mat\uptau \vec{\dot{x}} = - \vec{X}$)                      \\
$\nabla\phi$                         & $-\xi$                                &                            & $\mat\uptau$ (metric)                               \\ 
$\epsilon$      &     $-\sigma$ &    &     \\ \cline{1-2}
\multicolumn{2}{|c|}{1st law: $\omega=0$ on equilibrium manifolds} &                            &                          \\ \cline{1-2}
\multicolumn{2}{|c|}{2nd law: $\omega(\vec{t_p}) \leq 0 \iff$ CDI}           &                            &                          \\ \hline
\end{tabular}
\end{table}

The 3 GREs (1 per state variable) are now:

\begin{equation}
\tau_1 \dot{\phi} = - (Gg'(\phi)+[H_B(\mat{\epsilon})-H_A(\mat{\epsilon})]h'(\phi) + \pi)
\end{equation}

\begin{equation}
\tau_2 \nabla\dot{\phi} = - (\Gamma\nabla\phi - \mat\xi)
\end{equation}

\begin{equation}
\tau_3 \mat{\dot{\mat\epsilon}} = - (\overline{\mat{C}}(\phi)\mat{\epsilon} - \mat\sigma)
\end{equation}

With $\overline{\mat{C}}(\phi)= \mat{C}^A (1-h(\phi)) + \mat{C}^B h(\phi)$ the homogenized elastic tensor.\\

Note that we have 5 unknowns ($\phi$, $\mat\epsilon$, $\pi$, $\xi$, $\mat\sigma$) and 5 equations (3 GREs and the 2 momentum balances). Like before, we obtain the PFM by coupling the two first equations using the micro-force balance $\nabla.\xi + \pi = 0$:

\begin{equation}
\boxed{-\tau_2\Delta\dot{\phi} + \tau_1\dot{\phi} = \Gamma \Delta\phi - f(\phi,\mat\epsilon)}
\end{equation}

Where $f(\phi,\epsilon) = Gg'(\phi)+[H_B(\mat{\epsilon})-H_A(\mat{\epsilon})]h'(\phi)$ is the bulk energy term.

The obtained mechanical constitutive law is actually a viscoelastic-damage-like law:

\begin{equation}
\boxed{\mat\sigma = \overline{\mat{C}}(\phi)\mat{\epsilon} + \tau_3 \mat{\dot{\epsilon}}}
\end{equation}

We notice again that the non-equilibrium framework provides the new equations with a rate-dependency basically, i.e. a viscous term, compared with their conventional forms.  It can be seen macroscopically as the Voigt model for viscoelasticity, i.e. a Newtonian damper and Hookean elastic spring connected in parallel. More precisely, it can be seen as a microstructural/damage viscoelastic law.



The associated computation workflow is as follows:\\
\\
1) Initiate with boundary and initial conditions \\
2) Compute free energy $\Psi(\phi,\nabla\phi,\mat\epsilon)$ \\
3) Compute new $\phi$ with PFM equation \\
4) Deduce interpolation function $h(\phi)$ \\
5) Deduce new $\mat\epsilon$ from macro-force balance $\nabla.(\overline{\mat{C}}(\phi)\mat{\epsilon} + \tau_3 \mat{\dot{\epsilon}})=0$\\
6) Compute new $\Psi(\phi,\nabla\phi,\mat\epsilon)$ \\
(iterate from step 3)\\

Note again that the Voigt homogenization process is modified. Instead of imposing a pure interpolation of the stress \cite{Ammar2009} $\bar\sigma=h(\phi)\sigma_B + (1-h(\phi))\sigma_A$, the homogenized stress is naturally obtained from the mechanical GRE: $\bar\sigma=\overline{\mat{C}}(\phi)\mat{\epsilon} + \tau_3 \mat{\dot{\epsilon}} = h(\phi)\mat\sigma_B + (1-h(\phi))\mat\sigma_A + \tau_3 \mat{\dot{\epsilon}}$, with the additional viscous term.\\

The PFM can be naturally upscaled through the coupled mechanical GRE, which relates the homogenized stress and the homogenized strain. It yields a macroscopic viscoelastic-damage-like law, which actually acts as an upscaling of a potentially considered REV (representative elementary volume). It describes the upscaled mechanics without considering explicitly the interfaces anymore. The micro-scale or interfacial scale information is carried up through the homogenized mechanic tensor $\overline{\mat{C}}(\phi)$. The model thus shows off two coupled scales, the lower interfacial scale, where the PFM prevails, which we may call the microscale, and the upper non-interfacial scale, where the macro-force balance prevails $\nabla.\mat\sigma=\nabla.(\overline{\mat{C}}\mat{\epsilon} + \tau_3 \mat{\dot{\epsilon}}) = 0$,  which we may call the mesoscale. Then, the mechanics of a certain assemblage of REVs can be upscaled to a more engineering-like scale, which we may call the macroscale. We may conjecture that the mechanics of this macroscale will be defined by the average of the $\overline{\mat{C}}_{REVs}(\phi)$ for a given set of $REVs$.\\

We conclude the derivation of our model by writing the dimensionless form of the CPFM. Since the equation is in the current form homogeneous to a volumetric energy, we first divide by a characteristic energy, which we take as $G$. Then, normalizing the time and lengths respectively by $t_0$ and $l_0$, we get:

\begin{equation}
-\frac{\tau_2}{G}\frac{\Delta^*\dot{\phi}}{l_0^2t_0} + \frac{\tau_1}{G}\frac{\dot{\phi}^*}{t_0} = \frac{\Gamma}{G} \frac{\Delta^*\phi}{l_0^2} - \frac{f(\phi,\mat\epsilon)}{G}
\end{equation}

Where $*$ denotes the normalized derivatives, which we may as well drop in the following equations. We choose $t_0=\frac{\tau_1}{G}$ and $l_0$ as the problem's characteristic length scale in order to get:

\begin{equation}
\boxed{-\mu\Delta\dot{\phi} + \dot{\phi} = \alpha\Delta\phi - f_\chi(\phi,\mat\epsilon)}
\end{equation}

$\mu=\frac{\tau_2}{\tau_1 l_0^2}$ , 

With $\mu=\frac{\tau_2}{\tau_1 l_0^2}$ (phase-field viscosity), $\alpha=\frac{\Gamma}{G l_0^2}$ (dimensionless interfacial energy), $f_\chi(\phi,\mat\epsilon)=g'(\phi) + \chi(\mat\epsilon) h'(\phi)$ and $\chi(\mat\epsilon) = \frac{H_B(\mat{\epsilon})-H_A(\mat{\epsilon})}{G}$ (dimensionless input of energy, mechanical in this case).

We can thus see that the CPFM is driven by two dimensionless groups $\mu$ that we may call the phase-field viscosity and $\chi$ the bulk energy input, corresponding here to be the mechanical loading. While $\alpha$ is kept constant, $\chi$ will be shown (cf part \ref{LSA}) to destabilize the double-well stable configuration (and trigger the phase change) and $\mu$ to control the kinetics of the phase change (convergence to equilibrium).  This is corroborated numerically in \cite{Guevel2019b}.

\subsection{Extension to chemo-mechanical coupling}

Taking inspiration from \cite{Xu2011}, we extend our model to include chemical effects, namely the dissolution and precipitation. Still focusing on geomaterials, a natural application will be pressure solution creep \cite{Guevel2019b}.

Our motivation in extending the previous model is to broaden the instruments to make the most of PFM.  We have already added a complementary degree of freedom for the interface's movements by allowing the gradient of the order parameter to dissipate. Now we aim at using a counterpart of the natural phase change direction of PFM with mechanical loading, the production of weak phase A (cf part \ref{LSA}). Indeed PFM intrinsically prescribes the production of the least energetic phase, since the derivation is based on the minimization of the system's free energy, or more generally in our case to reach as fast as possible equilibrium. In the previous equations, the production of the weak phase A is favoured as it has the least (elastic) energy, with respect to the mechanical coupling. If we consider chemical reactions, the model should allow both creation of products and reactants, hence allow the creations of phase B as well. To better understand the phase change directionality, one can look at the tilt of the double well coupled with the bulk energy loading (cf part \ref{LSA}), or equivalently at the sign of the bulk energy. We thus have to add a bulk energy term that can counteract the mechanical energy. In the same manner the mechanical energy is defined, one can simply add the term $\beta c (1-h(\phi))$ in the free energy definition ($c$ is the solute concentration and $\beta$ a chemical coupling coefficient), so that the solute is only present in the pore phase. However, this form will be modified to guaranty mass conservation of the reactive species at stake, present either in the solid phase or either in the liquid phase (solute), under the form $\hat\beta(\phi)c = \tilde\beta(t)(b(t)+1-h(\phi))c$. In addition,  a source/sink term is to be included in the concentration equation. Similarly to the previous construction, the solute concentration is considered as a state variable, along with its gradient. We now consider the following expression for the equilibrium free energy:

\begin{equation}
\Psi(\phi,\nabla\phi,c\nabla c) = Gg(\phi) + \frac{1}{2}\mat\epsilon.(h(\phi)\mat{C}_B+(1-h(\phi))\mat{C}_A)\mat\epsilon +\hat\beta(\phi)c +\tau_{s}\dot{\phi} c + \frac{\Gamma}{2}||\nabla\phi||^2 + \frac{D}{2}||\nabla c||^2
\end{equation}

With $\tau_{s}$ the relaxation time for the solubility of the strong phase B.
In the same way the equation for the order parameter $\phi$ was derived, we get:

\begin{equation}
-\tau_4\Delta\dot{c} + \tau_3\dot{c} = D\Delta{c} - \tau_{s}\dot\phi - \hat\beta(\phi)
\end{equation}

Neglecting the dissipation of the interface's change of curvature ($\tau_5=0$), we recover a similar equation to \cite{Xu2011}. $\hat\beta$ is then determined to conserve the "mass", i.e. to have $\int_V (\dot\phi+\dot{c}) dV=0$. That is achieved by choosing:

\begin{equation}
\hat\beta(\phi) = \beta \left[1-h(\phi)- \overline{1-h(\phi)} \right]
\end{equation}

Where $\overline{1-h(\phi)}=\frac{\int_V(1-h(\phi))dV}{\int_V dV}$.

One can check that the mass balance is verified with null Neumann boundary condition for $c$ and the same coefficient for $\dot\phi$ as for $\dot{c}$, which will be obtained in the following dimensionless form:

\begin{equation}
\tau^*\dot{c} = D^*\Delta{c} - \tau^*\dot\phi - \hat\beta^*(\phi)
\end{equation}

With $\tau^*=\tau_4/\tau_1$, $\tau_{s}=\tau^*$, $\hat\beta^*(t)=\hat\beta/G$.\\

So that the final dimensionless system for CPFM with chemo-mechanical coupling reads:

\begin{equation} \label{chemo-mechanical system of equations}
\begin{cases} 
-\mu\Delta\dot{\phi} + \dot{\phi} = \alpha\Delta\phi - f(\phi,\mat\epsilon,c) \\ 
\tau^*\dot{c} = D^*\Delta{c} - \tau^*\dot\phi - \hat\beta^*(\phi)
\end{cases}
\end{equation}

With $f(\phi,\mat\epsilon,c)=g'(\phi)+(\chi(\mat\epsilon)-\beta^* c)h'(\phi)=g'(\phi)+\hat\chi(\mat\epsilon,c)h'(\phi)$. The sign of $\hat\chi(\mat\epsilon,c)$ will now govern the directionality of the phase changes, as described in part \ref{LSA} (cf fig.\ref{fig:double_well_tilting}).

\subsection{Reaction-diffusion systems} \label{Reaction-diffusion systems}

Given the broad development of the RDSs and their similarity with PFM, it is interesting to look at PFM under the perspective of RDSs. RDSs have been first introduced by Turing \cite{Turing1952} with application to morphogenesis and have remained mostly applied to biology. They consist of competing diffusive reactants, i.e. diffusing at different rates and reacting together. Patterns formation, or self-organization, corresponds to spatially inhomogeneous steady-states and can happen when the equilibria/steady-states are unstable, which can be triggered by mere fluctuations. This sudden deviation from the initial equilibrium state corresponds to a bifurcation. 

Using this analogy, our first equation for $\phi$ can be considered as the activation equation and the second one for $c$ as the inhibition equation. The main characteristics of RDSs can be retrieved. First, the diffusion kinetics of the activator and inhibitor should be significantly different; in our case, the solute concentration diffuses much faster than the physical interfaces governed by the order parameter. Second, the activator is auto-catalytic; the production of the weak phase A enhances stress/strain concentration and in return the latter enhances the former. Third, the activator catalyzes the production of inhibitor since the solute is directly produced from the dissolution of the solid phase B .

\section{Linear Stability Analysis} \label{LSA}


In this last part, let us perform a linear stability analysis of the model to show the change of stability with respect to the chemo-mechanical loading.

Let us consider a small perturbation around the steady states ${(\overline{\phi},\overline{c})}$, determined in \ref{Steady states}. We consider for simplicity a fixed strain state and a one-dimension problem. We thus write:

\begin{equation}
\begin{cases} 
\phi = \overline{\phi} + \tilde{\phi}\cos(kx)e^{\nu t}  \\ 
c = \overline{c} + \tilde{c}\cos(kx)e^{\nu t}
\end{cases}
\end{equation}

Where:

\begin{equation} \label{steady states equations}
\begin{cases}
\alpha\Delta\overline{\phi} - f(\overline{\phi},\overline{c}) = 0 \\
\Delta\overline{c} = 0
\end{cases}
\end{equation}

And:

\begin{equation}
\begin{cases} 
\tilde{\phi} \ll \overline{\phi}  \\ 
\tilde{c} \ll \overline{c}
\end{cases}
\end{equation}

$k$ denotes the wave number of the perturbation and $\nu$ the growth rate of the perturbation.

Let us linearize $f$ around the steady state $(\overline{\phi},\overline{c})$: 

\begin{equation}
f(\phi,c) = f(\overline{\phi},\overline{c}) + \frac{\partial{f}}{\partial{\phi}}(\overline{\phi},\overline{c})(\phi-\overline{\phi}) + \frac{\partial{f}}{\partial{c}}(\overline{\phi},\overline{c})(c-\overline{c}) + o(\phi-\overline{\phi},c-\overline{c})
\end{equation}

Injecting those expressions in the eq.\ref{chemo-mechanical system of equations} and noting $f_\phi = \frac{\partial{f}}{\partial{\phi}}(\overline{\phi},\overline{c})$ and $f_c = \frac{\partial{f}}{\partial{c}}(\overline{\phi},\overline{c})$ yields:

\begin{equation}
\begin{cases}
\mu \tilde{\phi} k^2 \nu \cos(kx) e^{\nu t} + \tilde{\phi}\nu \cos(kx) e^{\nu t} = \alpha(\frac{\partial{\overline{\phi}}}{\partial x^2} - \tilde{\phi}k^2 \cos(kx) e^{\nu t }) - f(\overline{\phi},\overline{c}) - f_\phi \tilde{\phi}\cos(kx)e^{\nu t} -f_c \tilde{c}\cos(kx)e^{\nu t} \\
\tau^*\tilde{c}\nu \cos(kx)e^{\nu t} = D^*(\frac{\partial{\overline{c}}}{\partial x^2} - \tilde{c} k^2 \cos(kx) e^{\nu t}) - \tau^* \tilde{\phi} \nu \cos(kx) e^{\nu t}+\hat\beta^*(\bar\phi,\bar{c})+\hat\beta^*_\phi \tilde\phi \cos(kx) e^{\nu t} +\hat\beta^*_c \tilde{c} \cos(kx) e^{\nu t}
\end{cases}
\end{equation}

Using the the steady state definition eq.\ref{steady states equations} and simplifying yields:

\begin{equation}
\nu \begin{pmatrix}
    1+\mu k^2       & 0  \\
   \tau^*      & \tau^*  \\
\end{pmatrix}
\begin{pmatrix}
 \tilde{\phi} \\ \tilde{c}  \\
\end{pmatrix} = 
\begin{pmatrix}
   -\alpha k^2 - f_\phi        & -f_c  \\
   -\beta^*h'(\bar\phi)      & -D^*k^2  \\
\end{pmatrix}
\begin{pmatrix}
 \tilde{\phi} \\ \tilde{c}  \\
\end{pmatrix}
\end{equation}

And further yielding the dispersion relation (linking $k$ and $\nu$, noting $\mat{M_k}$ the dispersion matrix):

\begin{equation}
\nu 
\begin{pmatrix}
 \tilde{\phi} \\ \tilde{c}  \\
\end{pmatrix} = 
\begin{pmatrix}
   -\frac{\alpha k^2 + f_\phi}{1+\mu k^2}        & 0  \\
   \frac{\beta^*h'(\bar\phi)}{1+\mu k^2}      & -\frac{D^*k^2}{\tau^*}  \\
\end{pmatrix}
\begin{pmatrix}
 \tilde{\phi} \\ \tilde{c}  \\
\end{pmatrix} = \mat{M_k}(\phi,c) \begin{pmatrix}
 \tilde{\phi} \\ \tilde{c}  \\
\end{pmatrix}
\end{equation}

Two necessary conditions for stability (i.e. ensuring eigenvalues with negative real parts) are $\det(\mat{M_k})=\frac{Dk^2}{\tau^*(1+\mu k^2)}(f_\phi+\alpha k^2) > 0$ (product of the two eigenvalues) and $\tr(\mat{M_k})=-\left(\frac{f_\phi+\alpha k^2}{1+\mu k^2} + \frac{Dk^2}{\tau} \right) < 0$ (sum of the two eigenvalues). Losing one of those two conditions can induce instability.



One can show that it is the same condition without the chemical coupling. The only difference is that $\hat\chi$ can become negative with chemical coupling.


Now let us evaluate this stability condition in the three possible steady states (derived in \ref{Steady states}). Note that $f_\phi=2(6\phi^2-6\phi(\hat\chi+1)+3\hat\chi+1)$.\\

For the steady state $\phi=0$ we have:

\begin{equation}
\begin{cases}
\det(\mat{M_k}(\phi=0))=\frac{Dk^2}{\tau^*(1+\mu k^2)}(6\hat\chi+2+\alpha k^2)  \\
tr(\mat{M_k}(\phi=0))=-\left(\frac{6\hat\chi+2+\alpha k^2}{1+\mu k^2} + \frac{Dk^2}{\tau} \right) < 0 
\end{cases}
\end{equation}

The determinant is positive iff $\hat\chi>-\hat\chi_0=-1/3-\alpha k^2/6$. Then if that is the case, the trace is negative. Therefore, the steady state $\phi=0$ is stable iff $\hat\chi>-\hat\chi_0$.\\

For the steady state $\phi=1$ we have:

\begin{equation}
\begin{cases}
\det(\mat{M_k}(\phi=0))=\frac{Dk^2}{\tau^*(1+\mu k^2)}(-6\hat\chi+2+\alpha k^2)  \\
tr(\mat{M_k}(\phi=0))=-\left(\frac{-6\hat\chi+2+\alpha k^2}{1+\mu k^2} + \frac{Dk^2}{\tau} \right) < 0 
\end{cases}
\end{equation}

The determinant is positive iff $\hat\chi<\hat\chi_0=1/3+\alpha k^2/6$. Then if that is the case, the trace is negative. Therefore, the steady state $\phi=1$ is stable iff $\hat\chi<\hat\chi_0$.\\

Finally, the third steady state $\phi=\frac{3\hat\chi+1}{2}$ iff $\hat\chi \in [-1/3,1/3]$ (to have $\phi \in [0,1]$). But then the determinant $\frac{Dk^2}{\tau^*(1+\mu k^2)}(9\hat\chi^2-1+\alpha k^2)$ is positive iff $|\hat\chi|>\sqrt{1-\alpha k^2}/3=\chi_1 \approx 1/3$ if $\alpha k^2 \ll 1$. Thus this steady state is unstable and can be disregarded if the problem's length scale $l_0$ is much larger than $\sqrt\alpha$ (since $k \propto 1/l_0$); as shown in \ref{LSA}, a system around the third (unstable) steady state will bifurcate to the first (lower horizontal branch) or second steady state (upper horizontal branch). We will assume that this is the case, i.e. that the problem's length scale is considerably larger than the interfaces characteristic width.\\

Gathering all the different cases, we can summarize the stability of the steady states in the following table:

\begin{table}[h!]
\centering
\caption{LSA summary}
\label{LSA summary}
\begin{tabular}{|c|c|c|c|}
\hline
\textbf{Regimes}                      & Chemical               & Rest                                    & Mechanical              \\ \hline
\textbf{Loading}                      & $\hat\chi<-\hat\chi_0$ & $\hat\chi \in [-\hat\chi_0,\hat\chi_0$] & $\hat\chi > \hat\chi_0$ \\ \hline
\textbf{$\phi=0$}                     & unstable               & \multicolumn{2}{c|}{stable}                                       \\ \hline
\textbf{$\phi=1$}                     & \multicolumn{2}{c|}{stable}                                      & unstable                \\ \hline
\textbf{$\phi=\frac{3\hat\chi+1}{2}$} & \multicolumn{3}{c|}{(conditionally) unstable}                                                       \\ \hline
\end{tabular}
\end{table}

\begin{figure}[h!]
\centering
  \includegraphics[width=.4\linewidth]{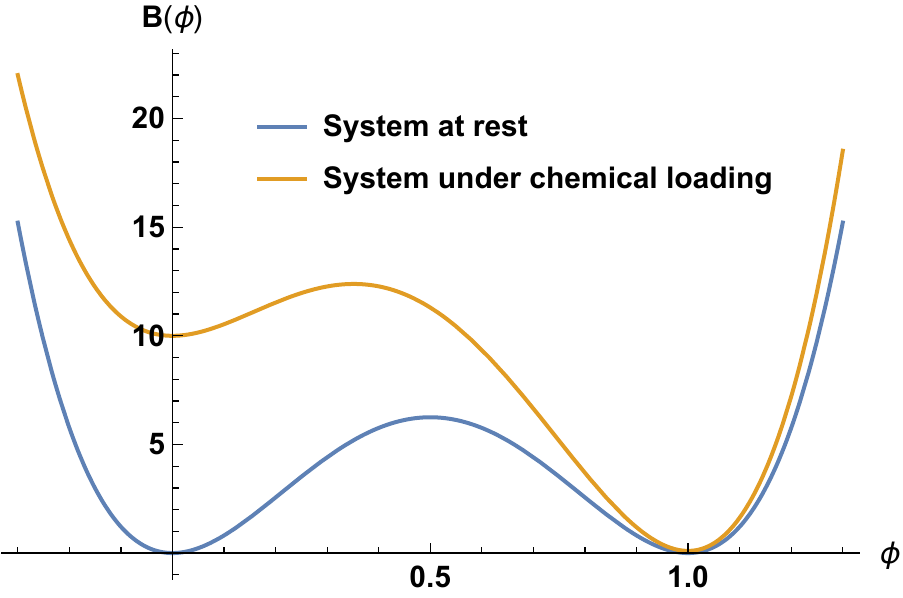}\hfill
  \includegraphics[width=.4\linewidth]{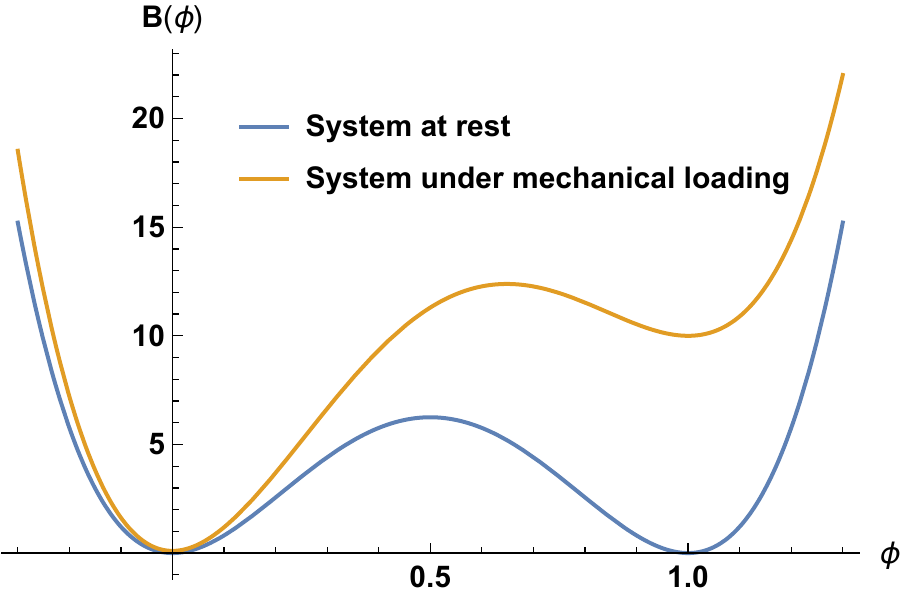}
\caption{Chemical and mechanical regimes destabilizing the system at rest, tilted double-well representation}
\label{fig:double_well_tilting}
\end{figure}

Thus, assuming as usually that the problem's length scale is much larger than the interface's one, the system possesses two possibly stable steady states $\phi=0$ and $\phi=1$. The third steady state $\phi=3\chi/2+1/2$ can be then disregarded as it is mostly unstable and would not appear durably in numerical simulations. There is a mechanical regime for $\hat\chi>\hat\chi_0$, where phase A (stable) is produced at the expanse of phase B (unstable) for a sufficient loading, and conversely for the chemical regime when $\hat\chi<-\hat\chi_0$. as represented on fig.\ref{fig:double_well_tilting}.\\

\begin{figure}[h!]
\centering
\includegraphics[scale=0.4]{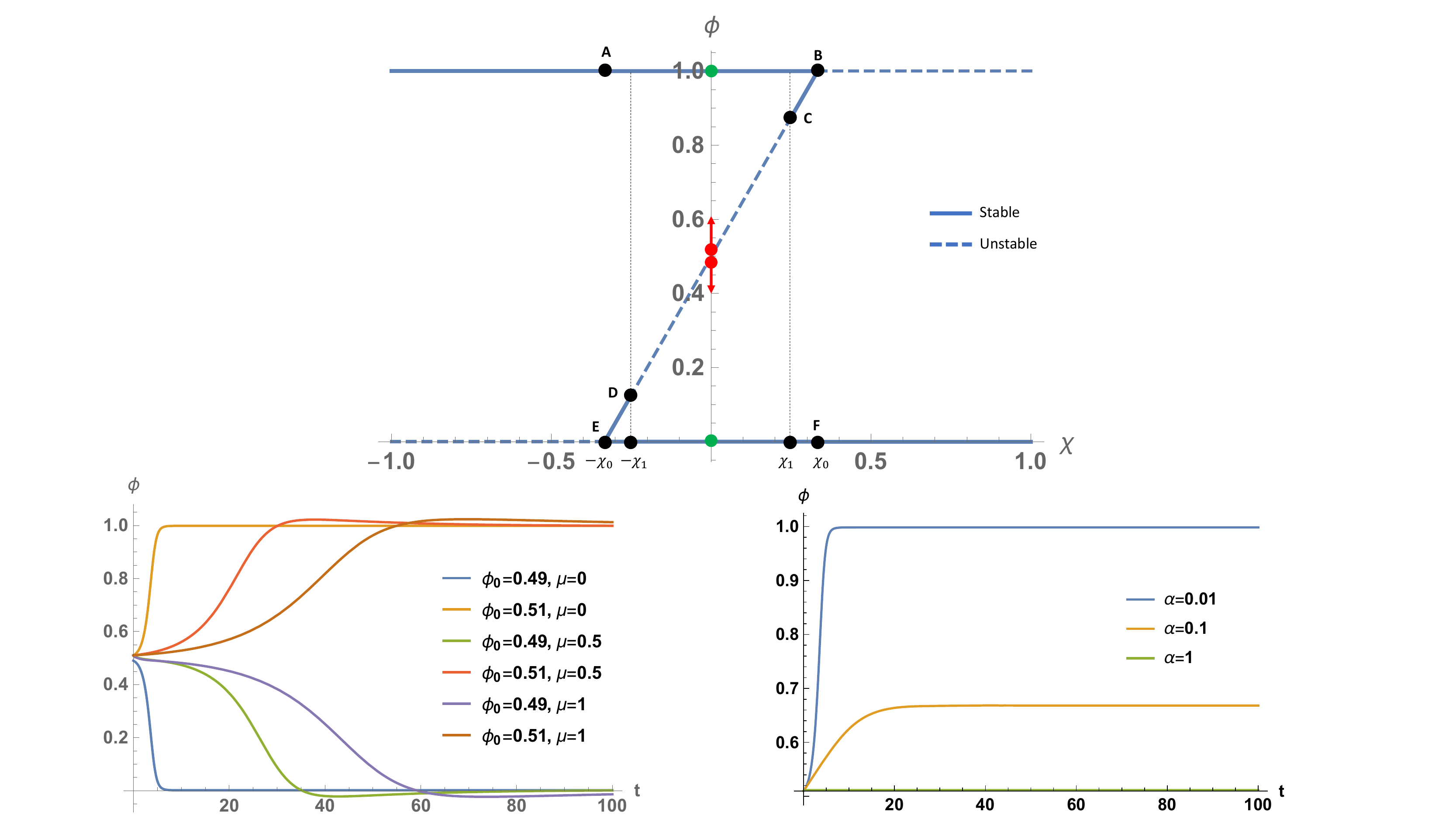}
\caption{Top: Z-curves with $\alpha k^2 <1$ ($\hat\chi_0=1/3+\alpha k^2/6$ and $\chi_1=\frac{\sqrt{1-\alpha k^2}}{3}$). Bottom left: transient convergence to the steady states $\phi=0$ and $\phi=1$ for $\alpha=0.01$, $\chi$=0, for different $\mu$. Bottom right: transient convergence to a possible third steady state for $\alpha$ large enough.}
\label{fig:lsa}
\end{figure}

We can summarize the previous stability results graphically as well with the following bifurcation curves (cf fig.\ref{fig:lsa}), which we call "Z-curves" (in reference to the "S-curves").

The two horizontal lines $[AB]$ and $[EF]$ represent the two stable states $\phi=0$ and $\phi=1$ respectively. The oblique line $[BE]$ corresponds to the third possible steady state $\phi=\frac{3\chi+1}{2}$ and can be unstable if $\alpha$ is large enough with respect to the problem's characteristic length; this case is usually to be avoided in PFM. Indeed, the smaller $\alpha$ the larger the unstable steady state's range $[CD]$; then two stable phases are to be predominantly observed. In Mathematica we can check that the third steady state is unstable for $\alpha$ small enough and without input of energy ($\chi=0$). 

For that we solve numerically the 1D problem:

\begin{equation}
-\mu\frac{\partial^3\phi}{\partial x ^2 \partial t}+\frac{\partial \phi}{\partial t} = \alpha \frac{\partial^2 \phi}{\partial x^2}  - 2\phi(1-3\phi +2\phi^2)
\end{equation}

We solve this equation associated with the initial condition $\phi(x,t=0)=\phi_0 \in \{0.49,0.51\}$ and the two boundary conditions $\phi(x=0,t)=\phi(x=1,t)=\phi_0$. We visualize the evolution of the point $x=1/2$. As expected, an initial state $\phi=0.51$ converges to the steady state $\phi=1$ whereas an initial state $\phi=0.49$ converges to the steady state $\phi=0$ (cf fig.\ref{fig:lsa} bottom left). Most interestingly, we can see that our new term characterized by $\mu$ delays the convergence to equilibrium. We can also check that the smaller $\alpha$ the closer to the sharp-interface problem , i.e. convergence to the steady states $\phi=0$ and $\phi=1$ (cf fig.\ref{fig:lsa} bottom right).

This is a preliminary appreciation of the effect of $\mu$ on a basic 1D problem without coupling. The convergence delay to equilibrium is translated into a delay of phase change, bringing rate-dependency intrinsically to the system. This corroborates the explicit relaxation equation GRE2 (cf eq.\ref{GRE_exponential_relaxation}), with affinities exponentially decreasing to 0 (equilibrium), characterized by the relaxation time (tensor) $\mat\uptau$. Recall that $\mu$ is the ratio of the PFM relaxation times (of $\phi$, $\tau_1$ and $\nabla\phi$, $\tau_2$). The Laplacian rate term does so by controlling the variations of the interfaces curvature, complementarily to the rate term controlling the interfaces' normal variations. More precisely, the interface (local mean) curvature (separating the phases $\phi=0$ and $\phi=1$) can be expressed as a function of the order parameter and approximated when $||\nabla\phi|| \ll 1$:

\begin{equation} \label{curvature}
\kappa \equiv -\nabla.\vec{n} \equiv -\nabla.\frac{\nabla\phi}{||\nabla\phi||} \approx \Delta\phi
\end{equation} 

With $\vec{n}$ the unit normal pointing towards the phase $\phi=1$; the curvature is thus positive if the interface curves towards the normal.

A (plane strain) 2D numerical study with chemo-mechanical problem is carried out in \cite{Guevel2019b} in FEM with various applications, particularly showing the interplay between the mechanical and chemical regimes, in the example of pressure solution.

\section{Conclusion}

We have set forth in the first part of this work the theoretical foundations of CPFM. This is an extended PFM based on a non-equilibrium thermodynamic framework, CT, incorporating for now chemo-mechanical coupling. We have first thoroughly explained our motivations to work with non-equilibrium thermodynamics and proposed a formal development based on contact geometry, whence the names of CT and CPFM we coin. As for the chemo-mechanical coupling, the mechanical effect is based on elasticity triggering the production of weak phase, similarly to dissolution. The chemical effect allows the opposite reaction, the production of strong phase, similarly to precipitation, in the zones away from the ones with large mechanical loading. The precise discrimination of the mechanical response within the system is ensured by the PFM capturing the actual interfaces. For this reason, we will apply this model in the second part \cite{Guevel2019b} to microstructures with complex geometries, those of geomaterials. This bidirectional interplay between an exothermic process (dissolution e.g.) and endothermic process (precipitation e.g.) can be more generally understood in context of reaction-diffusion systems.

The novelty of our extended PFM resides in the term $\mu \Delta\dot\phi$ ($\phi$ is the order parameter), added to the usual term $\dot\phi$. We claim that the latter characterizes the normal variations of the interfaces curvature and the former their change of orientations, i.e. tangential variations. Thus $\mu$, that we call PFM viscosity, quantifies the resistance for a rough geometry to smoothen. In that sense, $\mu$ encapsulates the kinetics of microstructural changes and could be described with the different activation energies of the catalizing/inhibiting effects associated to the main process, such as temperature, as shown off in the second part of this work. Three fundamental justifications for the presence of this term appear to us. From a kinematic point of view, since PFM is essentially modeling interfaces, i.e. a 2D object, the model should allow two degrees of freedom in the dynamics. From a fundamental thermodynamic point of view, since PFM is intrinsically a gradient theory, the gradient of the order parameter should be a full-fledged state variable of the system and hence should be allowed to dissipate. The latter argument is all the more significant as PFM is based on the MaxDP, at least in our construction, and as such all the dissipation should be available for maximization.

A LSA shows the phase change bidirectionality allowed by our chemo-mechanical coupling. Therein a simple 1D analysis sheds light on the role of $\mu$ as modeling the catalizing/inhibiting effects. The PFM viscosity indeed delays convergence to the steady state, i.e. equilibrium. Thus our CPFM allows to control both the phase change directionality as well as its kinetics. Those two main features physically represent production/consumption of strong phase (and conversely) and CI effects respectively. A relevant numerical application to pressure solution creep in geomaterials is carried out in the second part.


\appendix

 \section{Mathematical toolbox} \label{Mathematical toolbox}

We gather here some common notions of differential geometry from the literature, e.g. \cite{Sontz2015}. We do not intend to provide a thorough expos\'e of the discipline but rather the minimum viable knowledge useful for our model. In all the following parts, $M$ designates a smooth manifold. We recall that a manifold is a topological space that is a locally Euclidean Hausdorff space. That means that each point admits a neighborhood homeomorphic to the open unit ball in $\mathbb{R}^{m}$, where $m$ is the dimension of $M$. Such an homeomorphism (bicontinuous mapping) is called the chart of the manifold. Except in the first sub-part where $dim(M)=m$, we consider $dim(M)=2n+1$.

\subsection{Differential forms and associated operations}

Differential forms are crucial in differential geometry since they are the quantities assigning a measurement to the vector fields, which will give raise to the metric. It is important to know as well that differential forms act on tangential spaces. Those two characteristics make great sense of the importance of the differential forms in thermodynamics. Tangential spaces can be seen as dealing with the velocity of a curve at a certain point or in the context of thermodynamics, of a thermodynamic path, legitimized by the intrinsic metricity of the differential form. Indeed, as shown in our model's development, the Gibbs form governs the dynamics of the process onto a thermodynamic path. 

Let $k$ be a non-null natural integer. Differential k-forms are linear k-forms defined on tangent spaces. As such, differential forms are endowed with the exterior algebra structure of the linear forms and therefore with the associated exterior product.
The tangent space at a point $p \in M$ noted $T_pM$ contains all the vectors tangent to $M$ at $p$. The collection of all the tangent spaces is called the tangent bundle $TM$ ($TM = \bigcup_{p \in M} T_pM$). Finally, we define the $k^{th}$ power of the fiber product of $TM$, noted $T^{\diamond k}M$ as the space of all k-tuples of vectors tangent to M, i.e. from $TM \times ... \times TM$ (k times), at the same point of M.

 A differential k-form is a smooth map from $T^{\diamond k}M$ to $\mathbb{R}$ that is k-multilinear in each fiber of $T^{\diamond k}M$ and that is antisymmetric.  The differential k-forms on $M$ form a vector space noted $\Omega^k(M)$.\\
 
The main operations on a k-form are the wedge or exterior product ($\wedge$) and the exterior derivative ($d$). We restrict the definitions to the one-forms.

The wedge product of two one-forms $\alpha$ and $\beta$ in $\Omega^1(M)$ at a point $p \in M$ is the alternating bilinear two-form $(\alpha \wedge \beta)_p \in \Omega^2(M)$ defined by:

\begin{equation}
\forall (u,v) \in T_pM \times T_pM, \ (\alpha \wedge \beta)_p(u,v) = \alpha_p(u)\beta_p(v) - \alpha_p(v)\beta_p(u)
\end{equation}

As a consequence, $\alpha \wedge \beta = - \beta \wedge \alpha$ and $\alpha \wedge \alpha = 0$.

Writing the 1-form $\alpha$ in a local coordinates system $(w_1,...,w_m)$ as $\alpha = \sum_{i=1}^n a_i w_i$, the exterior derivative is defined as the two-form $d\alpha \in \Omega^2(M)$:

\begin{equation}
d\alpha = \sum_{i=1}^n da_i \wedge dw_i
\end{equation}

Inter alia, note that the exterior derivative of an exterior derivative is always null.\\

\subsection{Contact form}

Equipped with those two operations, we can now define a contact form, with which a manifold becomes a contact manifold.

A contact form $\alpha$ on the smooth manifold M of dimension $2n+1$ is a nowhere-zero differential one-form that is non-degenerate in the sense that $d\alpha$ is non-degenerate on $ker(\alpha)$. Equivalently, $\alpha$ is a contact form iff the $2n+1$ form $\alpha \wedge (d\alpha)^n$ is a volume form, i.e. $\alpha \wedge (d\alpha)^n \neq 0$ ($(d\alpha)^n=d\alpha \wedge ... \wedge d\alpha$ n times).

The hyperplane field (in $TM$) locally defined by $ker(\alpha)$ (of codimension 1 i.e. of dimension $2n$)  is called a contact structure. As per say, it is this contact hyperplane field that makes $M$ a contact manifold, more than $\alpha$ since the latter is not unique in the definition of the contact structure - it is defined up to multiplication by any nonvanishing function.

The contact condition $\alpha \wedge (d\alpha)^n \neq 0$ is equivalent to saying that $\alpha$ is maximally non-integrable, i.e. "as far from the integrality condition [$\alpha \wedge d\alpha=0$] as possible" \cite{Geiges2001}.

\subsection{Contact manifold}

The pair $(M,\alpha)$ - or more exactly $(M,ker(\alpha))$ (since $\alpha$ does not uniquely define the contact structure unlike $ker(\alpha$)) - is called a contact manifold if $\alpha$ is a contact form.

In a nutshell, a contact manifold is a smooth manifold associated with a hyperplane field that "maximally tangents" it - the contact structure "tangenting" the manifold cannot be "tangented" by another hypersurface.

\subsection{Legendre submanifold}

If the contact structure $ker(\alpha)$ is maximally non-integrable, some of its submanifolds can be integrable. Those integral submanifolds are called "isotropic submanifolds". The isotropic submanifolds of dimension $dim(ker(\alpha))/2=n$ are called "Legendre submanifolds".

A Legendre submanifold of $(M,\alpha)$ can be seen as the maximum solution of $\alpha=0$, since it can be shown that the maximal dimension of a submanifold of $ker(\alpha)$ is $dim(ker(\alpha)/2$) \cite{Geiges2001}.

\section{Tangential action of the Gibbs form} \label{Tangential action of the Gibbs form}

The tangent vector at a point $p$ of the TPS reads \cite{Haslach1997, Haslach2011}:

\begin{equation}
\vec{t_p} \equiv \frac{d\Psi^*}{dt} \frac{\partial}{\partial{\Psi^*}} + \sum_{i=1}^{n} \frac{dx_i}{dt} \frac{\partial}{\partial{x_i}} + \sum_{i=1}^{n} \frac{dy_i}{dt} \frac{\partial}{\partial{y_i}}
\end{equation}

The Gibbs form evaluated in $\vec{t_p}$ yields:

\begin{equation}
\begin{aligned}
\omega(\vec{t_p}) & =  \frac{d\Psi^*}{dt} - \sum_{i=1}^{n} x_i\frac{dy_i}{dt} \\
& = \frac{d\Psi}{dt} + \sum_{i=1}^{n} \frac{dx_i}{dt}y_i +\sum_{i=1}^{n} x_i\frac{dy_i}{dt} - \sum_{i=1}^{n} x_i\frac{dy_i}{dt} \\
& = \sum_{i=1}^{n} \frac{\partial\Psi}{\partial{x_i}}\frac{dx_i}{dt} + \sum_{i=1}^{n} y_i\frac{dx_i}{dt} \\
& = \sum_{i=1}^{n} (\frac{\partial\Psi}{\partial{x_i}}+y_i)\frac{dx_i}{dt} \\
& = \sum_{i=1}^{n} X_i\dot{x_i}
\end{aligned}
\end{equation}


\section{Conditioning bound lemma: corollary of the min-max theorem} \label{conditioning bound derivation}

Let be two vectors $\vec{u}$ and $\vec{v}$ collinear via the symmetric positive definite tensor $\mat\uptau$, i.e. $\vec{u}=\mat\uptau \vec{v}$. Then the angle $\theta$ between $\vec{u}$ and $\vec{v}$, reading $\theta_{(\vec{u},\vec{v})}=cos^{-1}(\frac{\vec{u}.\vec{v}}{||\vec{u}||||\vec{v}||})$, admits the upper bound $\theta_{\mat\uptau}$:

\begin{equation}
\theta_{(\vec{u},\vec{v})} \leq \theta_{\mat\uptau} = cos^{-1}(\frac{\lambda_{min}}{\lambda_{max}}) = cos^{-1}(cond(\mat\uptau)^{-1})
\end{equation}

With $\lambda_{min}$ and $\lambda_{max}$ respectively the minimum and maximum eigenvalues of $\mat\uptau$ (both strictly positive since $\mat\uptau$ is symmetric definite positive). $cond(\mat\uptau)$ is the so-called conditioning of $\mat\uptau$, ratio of its minimum and maximum eigenvalues.

\begin{proof}
Let us lower bound $cos(\theta_{(\vec{u},\vec{v})})$:

\begin{equation}
cos(\theta_{(\vec{u},\vec{v})}) = \frac{\mat\uptau\vec{v}.\vec{v}}{||\mat\uptau\vec{v}||||\vec{v}||} \geq \frac{\mat\uptau\vec{v}.\vec{v}}{||\mat\uptau||||\vec{v}||^2} = \frac{1}{\lambda_{max}} \frac{\mat\uptau\vec{v}.\vec{v}}{\vec{v}.\vec{v}} \geq \frac{\lambda_{min}}{\lambda_{max}}
\end{equation}

The first bounding results from the matrix norm inequality $||\mat\uptau\vec{v}|| \leq ||\mat\uptau||.||\vec{v}|| = \lambda_{max}.||\vec{v}||$ (the norm of a symmetric matrix is equal to its spectral radius). The second bounding corresponds to the lower bounding of the min-max theorem. Note that the resulting bounding is the tightest possible. 
Whence the upper bounding of $\theta_{(\vec{u},\vec{v})}$.

\end{proof}

\section{Steady states} \label{Steady states}

Let us first focus on the steady states for $\phi$, assuming $\mat\epsilon$ and $c$ fixed. In 1D they are the solutions of $\alpha \frac{\partial^2\phi}{\partial x^2}=4\phi^3 - 6(\hat\chi+1)\phi^2+(6\hat\chi+2)\phi$. As commonly done in PFM, we assume then RHS to be the leading order, when considering $\alpha$ (proportional to the interface width) much smaller than the other parameters. We are now looking for the solutions of the third-order polynomial $4\phi^3 - 6(\hat\chi+1)\phi^2+(6\hat\chi+2)\phi=0$. Two of the three solutions of this third-order polynomial are clearly 0 and 1 and we can rewrite the polynomial as $4\phi(\phi-1)(\phi-\frac{3\hat\chi+1}{2})$. The third root $\frac{3\hat\chi+1}{2}$ is physically admissible only if it is comprised between 0 and 1, i.e. if $\hat\chi \in [-1/3,1/3]$.

\section*{References}

\bibliography{library,library2}

\end{document}